\documentclass{aa}
\usepackage{txfonts}
\usepackage{epsfig}
\usepackage{subfigure}

\usepackage{times}
\usepackage{natbib}
\usepackage{rotating}
\bibpunct{(}{)}{;}{a}{}{,}

\begin{document}

\title{Tracing the evolution in the iron content of the ICM}

\author{I. Balestra\inst{1} \and P. Tozzi\inst{2,3} \and
S. Ettori\inst{4} \and P. Rosati\inst{5} \and S. Borgani\inst{3,6}
\and V. Mainieri\inst{1} C. Norman\inst{7} \and M. Viola\inst{6}}

\offprints{I. Balestra, \email{balestra@mpe.mpg.de}}

\institute{Max-Planck-Institut f\"ur Extraterrestrische Physik, Postfach 1312, 85741 Garching, Germany 
\and INAF, Osservatorio Astronomico di Trieste, via G.B. Tiepolo 11, I--34131, Trieste, Italy
\and INFN, National Institute for Nuclear Physics, Trieste, Italy 
\and INAF, Osservatorio Astronomico di Bologna, via Ranzani 1, I--40127, Bologna, Italy 
\and European Southern Observatory, Karl-Schwarzschild-Strasse 2, D-85748 Garching, Germany 
\and Dipartimento di Astronomia dell'Universit\`a di Trieste, via G.B. Tiepolo 11, I--34131, Trieste, Italy 
\and Department of Physics and Astronomy, Johns Hopkins University, Baltimore, MD 21218}

\date{Received 2006/ Accepted 2006}

\authorrunning{I. Balestra et al.}

\abstract {We present a Chandra analysis of the X-ray spectra of 56
clusters of galaxies at $z\,\ga 0.3$, which cover a temperature range of
$3\,\la kT\,\la 15$~keV.}  {Our analysis is aimed at measuring the
iron abundance in the intra-cluster medium (ICM) out to the highest
redshift probed to date.}  {We made use of combined spectral analysis
performed over five redshift bins at $0.3\la z \la 1.3$ to estimate
the average emission weighted iron abundance. We applied non-parametric
statistics to assess correlations between temperature, metallicity, and
redshift.}  {We find that the emission-weighted iron abundance
measured within $(0.15-0.3)\,R_{vir}$ in clusters below 5~keV is, on
average, a factor of $\sim2$ higher than in hotter clusters, following
$Z(T)\simeq 0.88\,T^{-0.47}\,Z_\odot$, which confirms the trend seen in
local samples.  We also find a constant average iron abundance
$Z_{Fe}\simeq 0.25\,Z_\odot$ as a function of redshift, but only for
clusters at $z\ga0.5$.  The emission-weighted iron abundance is
significantly higher ($Z_{Fe}\simeq0.4\,Z_\odot$) in the redshift
range $z\simeq0.3-0.5$, approaching the value measured locally in the
inner $0.15\,R_{vir}$ radii for a mix of cool-core and non cool-core
clusters in the redshift range $0.1<z<0.3$.  The decrease in
metallicity with redshift can be parametrized by a power law of the
form $\sim(1+z)^{-1.25}$.  We tested our results against selection
effects and the possible evolution in the occurrence of metallicity
and temperature gradients in our sample, and we do not find any evidence
of a significant bias associated to these effects.}  {The observed
evolution implies that the average iron content of the ICM at the present
epoch is a factor of $\sim2$ larger than at $z\simeq 1.2$.  We confirm
that the ICM is already significantly enriched ($Z_{Fe}\simeq0.25
\,Z_\odot$) at a look-back time of 9~Gyr. Our data provide significant
constraints on the time scales and physical processes that drive the
chemical enrichment of the ICM.}


\maketitle

\section{Introduction}

Clusters of galaxies are the largest virialized structures in the
Universe arising from the gravitational collapse of rare high peaks of
primordial density perturbations \citep[e.g.][]{pee93, col95, pea99, 
rbn02, v05}.  As a result of adiabatic compression and shocks
generated by supersonic motion during shell crossing and
virialization, a hot thin gas permeating the cluster's gravitational
potential well is formed. Typically this gas, which is enriched with
metals ejected form Supernovae (SNe) explosions through subsequent
episodes of star formation \citep[e.g.][]{mat88}, reaches temperatures
of several $10^7$ K and therefore emits mainly via thermal
bremsstrahlung in the X-rays. At such temperatures most of the
elements are either fully ionized or are in a high ionization
state. Strong emission lines may originate by collisional excitation
of K- and L-shell transitions in highly ionized elements, such as H-
and He-like iron, oxygen, silicon or sulfur. In the isothermal
approximation, the line intensities depend on the abundances of heavy
elements, while the continuum intensity is mainly due to hydrogen and
helium.  Therefore the equivalent width of a line, under the
reasonable assumption of collisional equilibrium, gives a direct
measurement of the abundance of the corresponding element.

The chemical enrichment of the intra-cluster medium (ICM) is an
unambiguous signature of star formation in cluster galaxies.  A
knowledge of the history of the ICM metal enrichment is also necessary
in understanding the mode and epoch of cluster formation and the
thermodynamic evolution of the cluster baryons in their hot and cold
phases. In this respect, measuring the properties of the ICM at high
redshift is important for constraining the physical processes involved in
the diffusion of energy and metals within clusters. In particular, the
chemical evolution of the ICM sets constraints on the SNe rate in the
cluster galaxies \citep[see][]{ett05}. The SNe explosions are, in fact,
the main contributor to the metal enrichment and are also expected to
provide a source of ICM heating \citep[e.g.][]{pip02, tor04}.

In addition, much emphasis has recently been given to the evolution of global
scaling relations for the ICM, such as the luminosity-temperature
relation and the entropy-temperature relation 
\citep{hol02,vik02,ett04,pra06,maugh06}.
Such scaling relations represent a further signature of
galaxy formation and super-massive black hole activity, whose relative 
contribution to the ICM energetic is under investigation.

\citet[ hereafter Paper~I]{toz03} measured for the first time the
average iron abundance in the ICM of hot clusters of galaxies out to
redshift $z\simeq1.3$. Their findings, which mostly probed the redshift
range $0.5<z<1.3$ (with 90\% of their sample in this range), suggested
that the mean iron content of the ICM is approximately constant with a
value $Z_{Fe}\simeq 0.3\,Z_\odot$\footnote{Solar abundance values are
set to those provided by \citet{and89}; in particular, the solar
abundance of iron atoms relative to hydrogen is $4.68\times 10^{-5}$. See
Sect.~3.1 for discussion of the solar iron abundance.} for clusters
with temperatures $kT>5$~keV. When comparing their results at high
redshift with the local values of the iron abundance, they referred to
the typical value $Z_{Fe}\simeq 0.3\,Z_\odot$ available from the
literature at that time \citep{ren97, deg01}.  Therefore, they concluded
that no evolution of iron abundance with redshift was found out to the
probed redshift.

More recently, the determination of an average local value of the ICM
iron content has become a harder task due to the different metal
distribution inside cool-core and non cool-core clusters \citep{deg04}. 
The former show a central peak of iron abundance with
$Z_{Fe}\simeq0.6-0.8\,Z_\odot$ and a plateau at $Z_{Fe}\simeq
0.3\,Z_\odot$ in the outer regions \citep[see][]{tam04, vik05}, 
and the latter a somewhat lower value 
($Z_{Fe}\simeq0.2-0.3\,Z_\odot$) constant with radius. The iron-rich 
cores typically have a size of 100~kpc \citep[see][]{deg04, vik05}. 
Therefore, it is important to take into account any effects due to
different physical apertures when comparing different samples from the
literature.

Studies of local cluster samples have also found an increase in the
iron abundance in clusters with temperatures $\la5$~keV
\citep[see][]{arn92, fin01, bag05}, whose physical interpretation is
still a matter of debate.

Single observations of high-z ($z>1$) clusters confirmed that
$Z_{Fe}\simeq 0.3\,Z_\odot$ or higher is common in the ICM
\citep{ros04, has04}. This implies that the last episode of star
formation in clusters of galaxies must have taken place at earlier
epochs in order to significantly enrich the diffuse medium with
metals. Therefore, the study of the iron abundance at high redshift is
expected to place strong constraints on the star formation history of
cluster galaxies and on its effects on the thermodynamics of the ICM.

Here we present a significantly improved analysis compared to our
previous work by substantially extending the sample (56 clusters
instead of the 19 presented in Paper~I) and by using the most recent 
{\em Chandra} calibrations in order to have an up-to-date data
reduction at the time of writing. This increase in statistics allows us
to investigate the relation between the iron abundance and global
temperature of the ICM at high redshift and to derive a more robust
measurement of the cosmic evolution of the average iron abundance in
the ICM, which is then compared with predictions from the cosmic star 
formation rate. 

The plan of the paper is as follows. In Sect.~2 we describe the data
reduction procedure. In Sect.~3 we describe our spectral analysis
(Sect.~3.1) and present the main results, subdivided as follows:
single source analysis (Sect.~3.2), correlation between iron
abundance and temperature (Sect.~3.3), evolution of the average iron
abundance as a function of redshift (Sect.~3.4), and proper comparison
with the {\em local} iron abundance (Sect.~3.5). In Sect.~4 we discuss
the implications of our findings and in Sect.~5 we summarize our
conclusions. In the Appendix, we describe in detail the spectral
simulations performed to investigate the possible spectral-fitting
biases due to unresolved temperature and metallicity gradients in our
sample. We adopt a cosmological model with $H_0=70$ km/s/Mpc,
$\Omega_M=0.3$, and $\Omega_\Lambda=0.7$ throughout.  Quoted confidence
intervals are $1\sigma$ unless otherwise stated.

\section{Sample selection and data reduction}

\subsection{Chandra data}

In Table~\ref{exposures}, we present the list of {\em Chandra} observations
analyzed in this paper. The selected sample consists of all the public
{\em Chandra} archived observations of clusters with $z\geq0.4$ as of
June 2004, including 9 clusters with $0.3< z < 0.4$. Some of them were
already presented in Paper~I. Data reduction is performed using the CIAO
3.2 software package with a recent version of the Calibration Database
(CALDB 3.0.0) including the correction for the degraded effective area
of ACIS-I chips due to material accumulated on the ACIS optical
blocking filter at the epoch of the observation. We also applied the
time-dependent gain
correction\footnote{http://asc.harvard.edu/ciao/threads/acistimegain/},
which is necessary to adjust the ``effective gains", which have been
drifting with time due to an increasing charge transfer inefficiency
(CTI).  Most of the observations were carried out with the ACIS-I
instrument, while for some clusters (see Table~\ref{exposures}) 
the Back Illuminated S3 chip of ACIS-S was also used.

\begin{table*}
\caption{\label{exposures}{\em Chandra} archive clusters sample.}
\centering
\begin{tabular}{l l l l l l l }
\hline\hline
Cluster & z & Obs. Id.$\mathrm{^a}$ & Exp. [ks]$\mathrm{^b}$ & Mode$\mathrm{^c}$ & 
$R_{ext}$ [$\arcsec$]$\mathrm{^d}$ & Net Counts$\mathrm{^e}$ \\
\hline
\object{MS $1008.1-1224$}        & 0.306 & 926       & 44    & I--V & 108.5 & 9260 \\
\object{MS $2137.3-2353$}        & 0.313 & 928       & 33    & S--V & 79.0 & 33000 \\ 
\object{Abell 1995}              & 0.319 & 906       & 56.4  & S--F & 103.0 & 30100 \\ 
\object{MACS J$0308.9+2645$}     & 0.324 & 3268      & 24.4  & I--V & 123.0 & 11200 \\
\object{ZwCl $1358.1+6245$}      & 0.328 & 516       & 48.3  & S--F & 88.5 & 19800 \\ 
\object{MACS J$0404.6+1109$}     & 0.355 & 3269      & 21.6  & I--V & 157.0 & 3100 \\ 
\object{RX J$0027.6+2616$}       & 0.367 & 3249      & 9.8   & I--V & 108.0 & 960 \\ 
\object{MACS J$1720.2+3536$}     & 0.391 & 3280      & 20.8  & I--V & 103.0 & 6670 \\ 
\object{ZwCl $0024.0+1652$}      & 0.395 & 929       & 39.5  & S--F & 64.0 & 3150 \\ 
\object{V $1416+4446$}           & 0.400 & 541       & 31.0  & I--V & 73.8 & 2130 \\ 
\object{MACS J$0159.8-0849$}     & 0.405 & 3265      & 17.6  & I--V & 118.0 & 8100 \\ 
\object{MACS J$2228.5+2036$}     & 0.412 & 3285      & 20    & I--V & 137.7 & 6070 \\ 
\object{MS $0302.7+1658$}        & 0.424 & 525       & 10.0  & I--V & 59.0 & 635 \\
\object{MS $1621.5+2640$}        & 0.426 & 546       & 30.0  & I--F & 118.0 & 3280 \\ 
\object{MACS J$0417.5-1154$}     & 0.440 & 3270      & 12    & I--V & 138.0 & 7400 \\ 
\object{MACS J$1206.2-0847$}     & 0.440 & 3277      & 23    & I--V & 138.0 & 11720 \\ 
\object{RX J$1347.5-1145$}       & 0.451 & 3592      & 57.5  & I--V & 128.0 & 62700 \\ 
\object{V $1701+6414$}           & 0.453 & 547       & 49.0  & I--V & 64.0 & 2745 \\ 
\object{CL $1641+4001$}          & 0.464 & 3575      & 45.0  & I--V & 49.0 & 1040 \\ 
\object{MACS J$1621.4+3810$}     & 0.465 & 3254      & 9.7   & I--V & 78.0 & 1600 \\ 
\object{MACS J$1824.3+4309$}     & 0.487 & 3255      & 14.8  & I--V & 84.0 & 530 \\ 
\object{MACS J$1311.0-0311$}     & 0.492 & 3258      & 14.8  & I--V & 79.0 & 2100 \\
\object{V $1525+0958$}           & 0.516 & 1664      & 50    & I--V & 79.0 & 2100 \\ 
\object{MS $0451.6-0305$}        & 0.539 & 529,902   & 56    & I/S--V & 98.4 & 16850 \\ 
\object{MS $0015.9+1609$}        & 0.541 & 520       & 67.0  & I--V & 98.4 & 16200 \\ 
\object{MACS J$1149.5+2223$}     & 0.544 & 1656,3589 & 38    & I--V & 148.0 & 9400 \\ 
\object{MACS J$1423.8+2404$}     & 0.545 & 1657      & 18.5  & I--V & 79.0 & 3600 \\ 
\object{MACS J$0717.5+3745$}     & 0.548 & 1655,4200 & 78    & I--V & 144.0 & 29000 \\ 
\object{V $1121+2327$}           & 0.562 & 1660      & 70.0  & I--V & 69.0 & 2050 \\ 
\object{SC $1120-1202$}          & 0.562 & 3235      & 68    & I--V & 49.0 & 730 \\ 
\object{RX J$0848.7+4456$}       & 0.570 & 927,1708  & 184.5 & I--V & 30.0 & 850 \\ 
\object{MACS J$2129.4-0741$}     & 0.570 & 3199      & 17.6  & I--V & 98.0 & 3000 \\
\object{MS $2053.7-0449$}        & 0.583 & 551,1667  & 88    & I--V & 54.1 & 2150 \\ 
\object{MACS J$0647.7+7015$}     & 0.584 & 3196      & 19.2  & I--V & 88.5 & 3170 \\ 
\object{RX J$0956.0+4107$}       & 0.587 & 5294      & 17.2  & I--V & 64.0 & 500 \\ 
\object{CL $0542.8-4100$}        & 0.634 & 914       & 50    & I--F & 78.7 & 2220 \\ 
\object{RCS J$1419.2+5326$}      & 0.640 & 3240      & 9.7   & S--V & 44.0 & 470 \\ 
\object{MACS J$0744.9+3927$}     & 0.686 & 3197,3585 & 40    & I--V & 98.0 & 6100 \\ 
\object{RX J$1221.4+4918$}       & 0.700 & 1662      & 78    & I--V & 78.7 & 2900 \\ 
\object{RX J$1113.1-2615$}       & 0.730 & 915       & 103   & I--F & 39.4 & 1200 \\ 
\object{RX J$2302.8+0844$}       & 0.734 & 918       & 108   & I--F & 54.0 & 1600 \\
\object{MS $1137.5+6624$}        & 0.782 & 536       & 117   & I--V & 49.2 & 4150 \\ 
\object{RX J$1317.4+2911$}       & 0.805 & 2228      & 110.5 & I--V & 24.5 & 240 \\ 
\object{RX J$1350.0+6007$}       & 0.810 & 2229      & 58    & I--V & 64.0 & 750 \\ 
\object{RX J$1716.4+6708$}       & 0.813 & 548       & 51    & I--F & 54.0 & 1520 \\
\object{RX J$0152.7-1357$} S     & 0.828 & 913       & 36    & I--F & 52.7 & 570 \\ 
\object{MS $1054.4-0321$}        & 0.832 & 512       & 80    & S--F & 78.7 & 10000 \\ 
RX J$0152.7-1357$ N              & 0.835 & 913       & 36    & I--F & 58.0 & 830 \\ 
\object{1WGA J$1226.9+3332$}     & 0.890 & 932,3180  & 9.5   & S--V & 64.0 & 2400 \\
\object{CL $1415.1+3612$}        & 1.030 & 4163      & 89    & I--V & 39.4 & 1320 \\ 
\object{RDCS J$0910+5422$}       & 1.106 & 2227,2452 & 170   & I--V & 24.6 & 440 \\ 
\object{RX J$1053.7+5735$} E$^*$ & 1.134 & 4936      & 94    & S--V & 28.2 & 300 \\
RX J$1053.7+5735$ W$^*$          & 1.134 & 4936      & 94    & S--V & 28.2 & 450 \\
\object{RDCS J$1252-2927$}$^*$   & 1.235 & 4198,4403 & 188.4 & I--V & 34.5 & 850 \\ 
\object{RDCS J$0849+4452$}$^*$   & 1.261 & 927,1708  & 184.5 & I--V & 23.6 & 360 \\ 
\object{RDCS J$0848+4453$}       & 1.273 & 927,1708  & 184.5 & I--V & 19.7 & 130 \\ 
\hline
\end{tabular}
\begin{list}{}{}
\item[Notes:] $\mathrm{^a}$ observation identification number; $\mathrm{^b}$ effective 
exposure time after removal of high background intervals; $\mathrm{^c}$ detector 
(ACIS-I or -S) and telemetry (FAINT or VFAINT) used; $\mathrm{^d}$ extraction radius; 
$\mathrm{^e}$ number of net detected counts in the $0.3-10$~keV band; $^*$ denotes 
clusters for which we also use XMM-{\em Newton} observations (see Table~\ref{xmmexp}).
\end{list}
\end{table*}

We started to process data from the level=1 event file.  For
observations taken in the VFAINT mode, we run the tool {\tt
acis\_process\_events} to flag probable background events using all
the information of the pulse heights in a $5\!\times \!5$ event island
(as opposed to a $3\!\times \!3$ event island recorded in the FAINT
mode) to help in distinguishing between genuine X-ray events and 
artificial events that are most likely associated with cosmic rays. With this
procedure, the ACIS particle background can be significantly reduced
compared to the standard grade selection\footnote
{http://asc.harvard.edu/cal/Links/Acis/acis/Cal\_prods/vfbkgrnd/}.
Real X-ray photons are hardly affected by such cleaning
(only less than 2\% of them are rejected, independent of the energy
band, provided there is no pileup). We also applied the CTI
correction\footnote{http://cxc.harvard.edu/
ciao/threads/acisapplycti/} to the observations taken when the
temperature of the focal plane was 153 K. This procedure allows us to
recover the original spectral resolution that is partially lost because of the
CTI. The correction applies only to ACIS-I chips, since the ACIS-S3
did not suffer from radiation damage.

For data taken in the FAINT mode we ran the tool 
{\tt acis\_process\_events} only to apply the CTI and the time-dependent
gain correction. From this point on, the reduction was similar for
both the FAINT and the VFAINT exposures. The data were filtered to
include only the standard event grades 0, 2, 3, 4, and 6. We checked
visually for hot columns left from the standard cleaning. Only in a few cases
did hot columns have to be removed by hand. We identify the
flickering pixels as the pixels with more than two events contiguous
in time, where a single time interval was set to 3.3~s. For exposures
taken in VFAINT mode, there were practically no flickering pixels left
after filtering out ``bad'' events. We finally filtered time intervals
with high background by performing a $3\sigma$ clipping of the
background level using the script {\tt
analyze\_ltcrv}\footnote{http://
cxc.harvard.edu/ciao/threads/filter\_ltcrv/}. Removed time intervals
always amount to less than 5\% of the nominal exposure time for
ACIS-I chips. Some ACIS-I observations show large flares on the
ACIS-S3 chip (which is on by default but not used in the data
analysis), but the corresponding time intervals are not removed since
the flares do not affect the ACIS-I chips. In any case, our spectral
analysis is not strongly affected by residual flares, since we always
compute the background from source-free regions around the clusters
from the same observation (see below), thus taking into account any
possible spectral distortion of the background itself induced by the
flares.

\begin{figure}
\centering
\includegraphics[width=8.0 cm, angle=0]{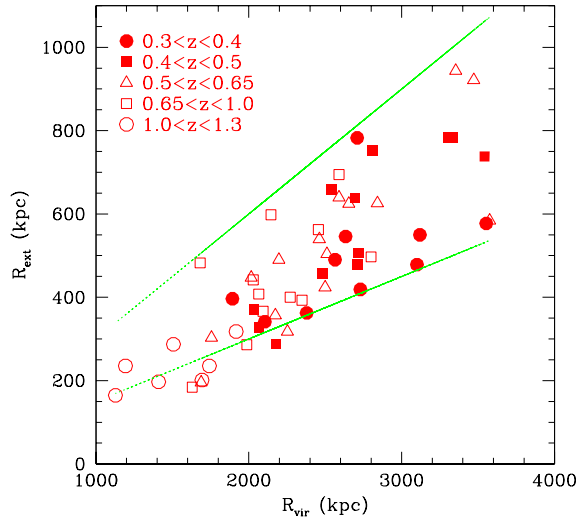}
\caption{Extraction radius ($R_{ext}$) versus virial radius
($R_{vir}$) for the whole sample. Lower and upper lines show
$R_{ext}=0.15\,R_{vir}$ and $R_{ext}=0.3\,R_{vir}$, respectively.}
\label{radius}
\end{figure}

As in Paper~I, we performed a spectral analysis extracting the spectrum
of each source from a region defined in order to maximize the
signal-to-noise ratio (S/N, see Appendix A.1 for details on how this
is computed). This choice of the extraction region allows the global
properties of the clusters to be measured using the majority of the
signal. This strategy is optimized for the highest redshift objects,
and it is homogeneously adopted for the whole sample. The region of
maximum S/N is obtained through the following procedure: the spectrum
of each source is extracted from a circular region around the centroid
of the photon distribution. For a given radius, we find the center of
the region that includes the maximum number of net counts in the
$0.5-5$~keV band, where the bulk of the source counts are
detected. Then, we compute the S/N, repeating this procedure for
several radii. Finally we choose the extraction radius $R_{ext}$,
defined as the radius for which the S/N is maximum. As shown in 
Fig.~\ref{radius}, in most cases $R_{ext}$ is between 0.15 and 0.3 times
the virial radius $R_{vir}$, estimated, after \citet{evr96}, as
\begin{equation}
R_{vir}=3.95\, \left( \frac{\mathit{T_{vir}}}{10\, \mathrm{keV}}\right) 
^{\frac{1}{2}}\, \mathit{F(z)} \, \, \mathrm{Mpc} \, .
\label{1} 
\end{equation}
Here the virial temperature $T_{vir}$ is approximated with the
spectral temperature $T_{spec}$ measured within $R_{ext}$, and
\begin{equation}
F(z) = (\Delta(z)/\Delta_0)^{-1/6}\, 
[\Omega_0\, (1+z)^3 + 1 - \Omega_0]^{-1/2} \, ,
\end{equation}
where $\Delta (z)$ is the density contrast of the virialized halo with
respect to the critical density. For simplicity we assume a
constant value here for ($\Delta(z)/\Delta_0)^{-1/6} $
\citep[see][]{bry98}. From Fig.~\ref{radius} we note that the
typical extraction region, depending both on the redshift through the
surface brightness dimming and on the brightness distribution of each
source, does not show a clear trend with redshift, with the exception
of the highest-z bin where $R_{ext}\leq 0.15\,R_{vir}$. The fraction
of net counts included in the extraction region always amounts to
$80-90$\% of the total detected for each cluster. We also note that
$R_{ext}$ is roughly 3 times the core radius measured with a beta model 
(see \citealt{ett03} for the spatial analysis of a subsample of 
our clusters).

For each cluster, we used the events included in each of the extraction
regions defined above to produce a spectrum (pha) file. The background
is always obtained from empty regions of the chip in which the source
is located. This is possible since all sources have an extension of
less than 3 arcmin, as opposed to the 8 arcmin size of the ACIS-I/-S 
chips. The background file is scaled to the source file
by the ratio of the geometrical area. The background
regions should partially overlap with the outer virialized regions of the
clusters. However, the cluster emission from these regions is
negligible compared to the instrumental background and does not
affect our results. Our background subtraction procedure, on the other
hand, has the advantage of providing the best estimate of the
background for that specific observation. By comparing the count rate in the source
and in the background at energies higher than 8~keV, we finally checked that
variations in the background intensity across the chip did not affect
the background subtraction, where the signal
from the clusters is null. The response matrices and the ancillary
response matrices of each spectrum were computed respectively with {\tt
mkacisrmf} and {\tt mkwarf} for the same regions from which the
spectra were extracted. For those observations for which the CTI
correction cannot be applied (when the temperature of the detector is
larger than $153$~K), we used {\tt acisspec} instead.

\subsection{XMM-Newton data}

As in Paper~I, we used the XMM-{\em Newton} data to boost the S/N only for
the most distant clusters in our current sample, namely the clusters
at $z>1$. In Table~\ref{xmmexp} we list the four XMM-{\em Newton}
observations of high redshift ($z>1$) clusters included in our
analysis. For each observation we used both the European Photon Imaging
Camera (EPIC) PN and the two MOS detectors. The XMM-{\em Newton} observations
and data reduction relative to RDCS~J0849 and the two clumps of RX~J1053
have already been presented in Paper~I, while those relative to
RDCS~J1252 are described in \citet{ros04}.

\begin{table*}
\caption{Additional XMM-{\em Newton} observations at $z>1$.}
\label{xmmexp}
\centering
\begin{tabular}{l l l l l l }
\hline\hline
Cluster & z & Exp. [ks]$\mathrm{^a}$ & Detector$\mathrm{^b}$ & 
$R_{ext}$ [$\arcsec$]$\mathrm{^c}$ & Net Counts$\mathrm{^d}$ \\
\hline 
RX J$105346.6+573517$ E & 1.134 & 94.5  & PN+2MOS & 32   & 708 \\ 
RX J$105346.6+573517$ W & 1.134 & 94.5  & PN+2MOS & 32   & 875 \\ 
RDCS J$1252-2927$       & 1.235 & 65.0  & PN+2MOS & 34.5 & 1570 \\ 
RDCS J$0849+4452$       & 1.261 & 112.0 & PN+2MOS & 29.5 & 630 \\
\hline
\end{tabular}
\begin{list}{}{}
\item[Notes:] $\mathrm{^a}$ effective exposure time after removal of high 
background intervals; $\mathrm{^b}$ detectors used; $\mathrm{^c}$ extraction 
radius; $\mathrm{^d}$ number of net detected counts in the $0.3-10$ keV band.
\end{list}
\end{table*}

\section{Results}

In this section we present the main results of our analysis. The section is
subdivided into five subsections. In Sect.~3.1 we provide a general description 
of our spectral analysis and a comparison with the previous results obtained 
in Paper~I. In Sect.~3.2 we describe the single source analysis and the main 
properties of the sample. In Sect.~3.3 the correlation between iron abundance 
and temperature is discussed. In Sect.~3.4 we present our results on the 
evolution of the average iron abundance as a function of redshift obtained 
through two independent methods ({\em combined fits} and {\em weighted means}). 
Finally, in Sect.~3.5 we present a comparison with the {\em local} iron 
abundance of the ICM.

\subsection{Spectral analysis}

The spectra were analyzed with XSPEC v11.3.1 \citep{arn96} and fitted
with a single-temperature {\tt mekal} model \citep{kaa92, lie95} in
which the ratio between the elements was fixed to the solar value as in
\citet{and89}. These values for the solar metallicities have more
recently been superseded by the new values by \citet{gre98} and
\citet{asp05}, who introduced a 0.676 and 0.60 times lower iron solar
abundance, respectively (photometric value). However, we prefer to
report iron abundances in units of solar abundances by \citet{and89}
since most of the literature still refers to them.  We also performed
the fits using solar abundances by \citet{asp05}.  The iron abundances
in these units (reported in the fifth column of Table~3) can be
obtained with an accuracy of about 10\% simply by rescaling the values 
measured in solar units by \citet{and89} by a factor of 1.6. This
shows that we are not affected by the presence of metals other than
iron.  Finally, we model Galactic absorption with {\tt tbabs}
\citep{wil00}.

It has recently been shown that a methylene layer on the {\em Chandra}
mirrors increases the effective area at energies higher than 2 keV
\citep{mar04}\footnote{
http://cxc.harvard.edu/ccw/proceedings/03\_proc/presentations/
marshall2}.  This has a small effect on the total measured fluxes, but
it may be non-negligible on the spectral parameters (i.e., it may
artificially reduce the temperatures). In order to correct for it, we
introduced a ``positive absorption edge'' (XSPEC model {\tt edge})
in the fitting model at 2.07 keV with $\tau =-0.15$ \citep{vik05}.

The fits were performed over the energy range $0.6-8.0$ keV. Due to 
uncertainties in ACIS calibration below 0.6 keV, we excluded less
energetic photons from the spectral analysis in order to avoid
systematic bias. The effective cut at high energies is generally lower
than $7-8$~keV, since the S/N for a thermal spectrum rapidly decreases
above 5 keV.

The free parameters in our spectral fits are temperature, metallicity,
and normalization. Local absorption is fixed to the Galactic neutral
hydrogen column density ($N_H$ in Table~\ref{results}), as obtained
from radio data \citep{dic90}, and the redshift to the value measured
from optical spectroscopy ($z$ in Table~\ref{results}).  We used Cash
statistics applied to the source plus
background\footnote{http://heasarc.gsfc.nasa.gov/docs/xanadu/xspec/manual/
XSappendixCash.html}, which is preferable for low S/N spectra
\citep{nou89}.

\begin{figure}
\centering
\includegraphics[width=7.0 cm, angle=0]{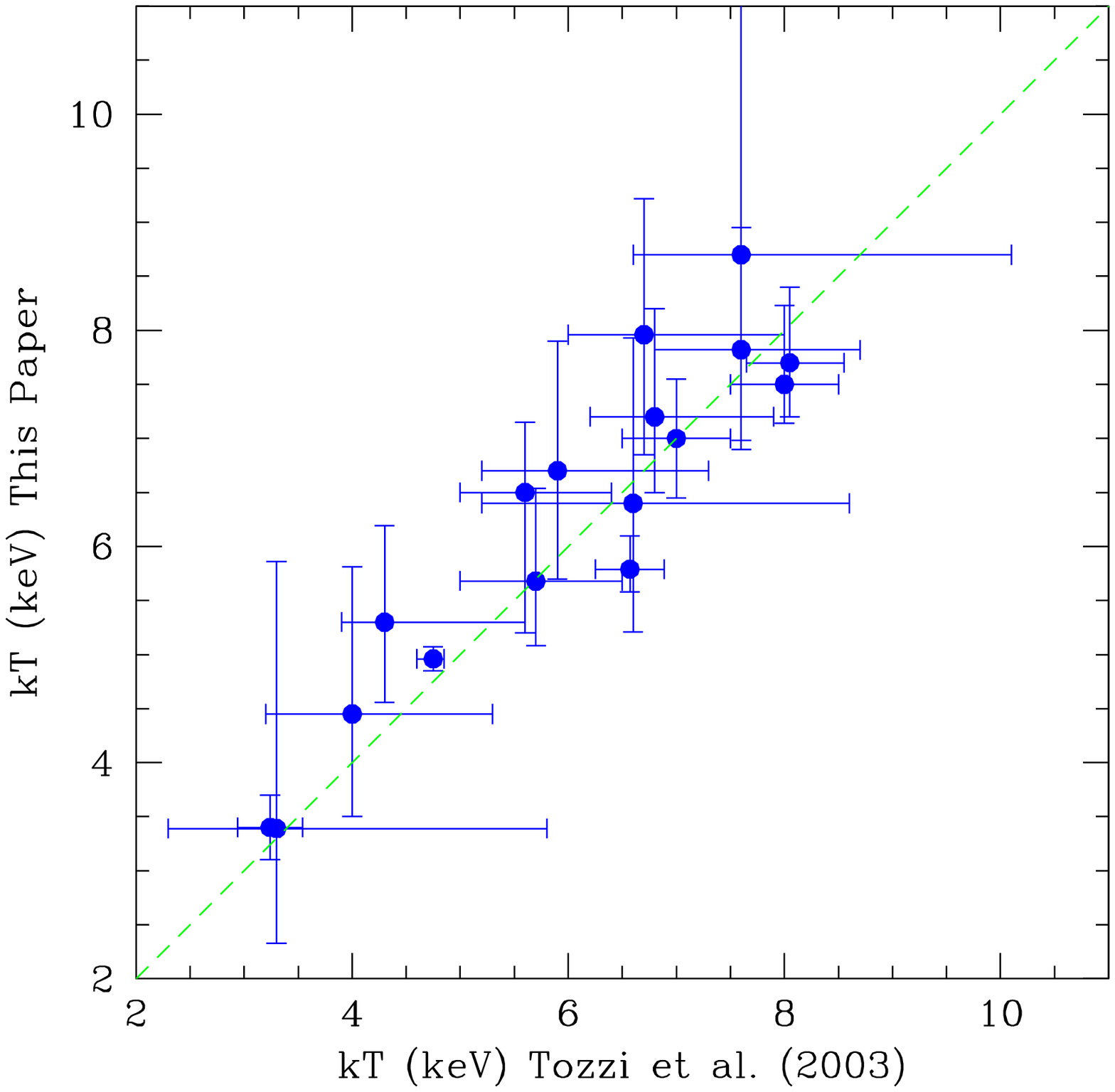}
\includegraphics[width=7.0 cm, angle=0]{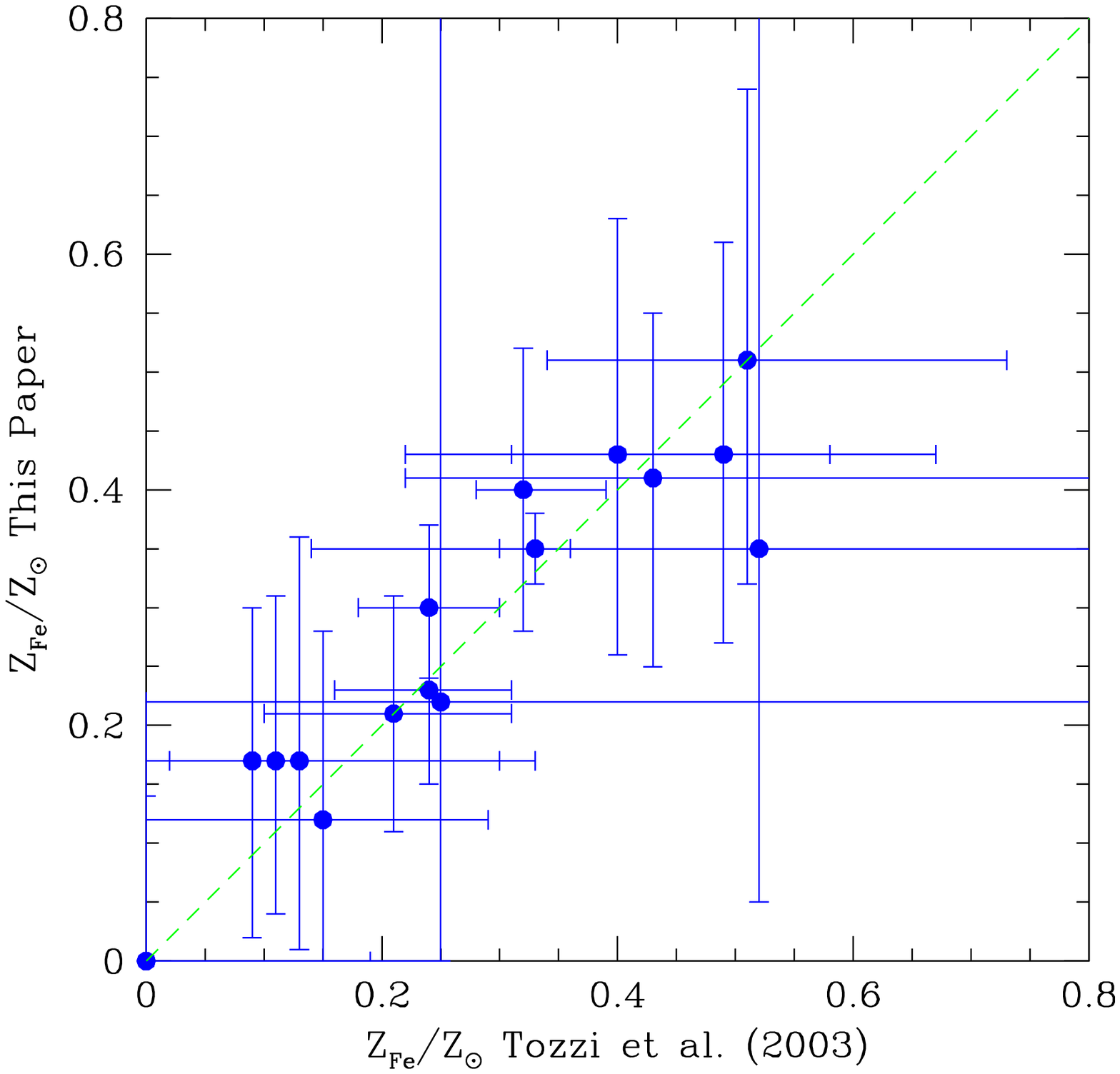}
\caption{Comparison between the temperature ({\em upper panel}) and
iron abundance ({\em lower panel}) values measured in Paper~I, and
those measured in this work after the most recent {\em Chandra}
calibrations have been applied. Dashed lines show the locus of equal
temperature and abundance values.}
\label{compare}
\end{figure}

\begin{figure}
\centering
\includegraphics[width=8.0 cm, angle=0]{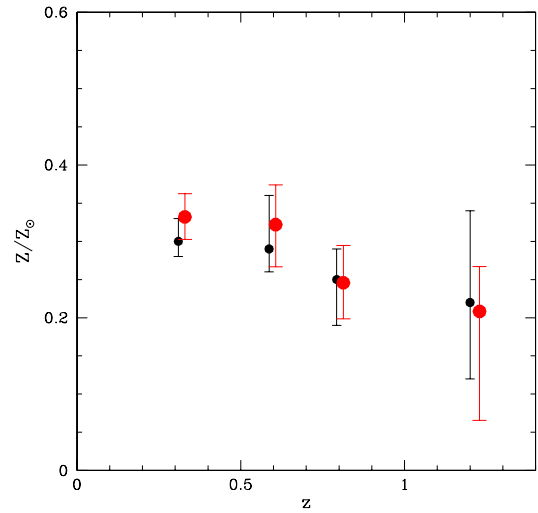}
\caption{Average iron abundance in different redshift bins computed
for the same sample of clusters analyzed in Paper~I using updated
calibrations (red circles), compared with the previous results (black
circles). The plot shows that the new calibrations and data reduction
have a negligible effect on our results. Only clusters with $kT>5$~keV
are considered.}
\label{newcal}
\end{figure}

Eventhough our analysis procedure is similar to the one used in Paper~I, 
new calibrations (including the treatment of the positive edge at
2.07~keV) may cause some differences in the new temperature and iron
abundance values compared to the previous analysis. To investigate
such differences, we show in Fig.~\ref{compare} the temperatures and
iron abundances published in Paper~I, plotted against the new values
(presented in this paper) for the 17 clusters observed with {\em
Chandra} in both samples, plus RDCS~1252 \citep{ros04}. The new
best-fit temperatures (upper panel) seem to be slightly higher than
in Paper~I, while iron abundances (lower panel) are much less affected
by the new calibrations. As a further check we recomputed the average
values of the iron abundance in the same redshift bins used in Paper~I
and we find consistent results (see Fig.~\ref{newcal}). The only
noticeable difference is the hint of a decrease with $z$, well
below the $2\sigma$ confidence level in Paper~I, which is now slightly
enhanced. Therefore, the most significant improvements with respect to
Paper~I are due to the new clusters included in the current sample. In
particular, we recall that the $z\simeq1.2$ point was entirely
dominated by the XMM data on RX~J1053, since in Paper~I there was no
statistically significant detection of the iron line at $z>1$ from
{\em Chandra} data only. The situation at $z>1$ has improved thanks to
the {\em Chandra} observations of CL~1415 and RDCS~1252. Finally, the
statistics of the current sample mostly improved in the redshift range
$0.3 < z < 0.6$.

\subsection{Single source analysis}

\begin{figure}
\centering
\includegraphics[width=8.0 cm, angle=0]{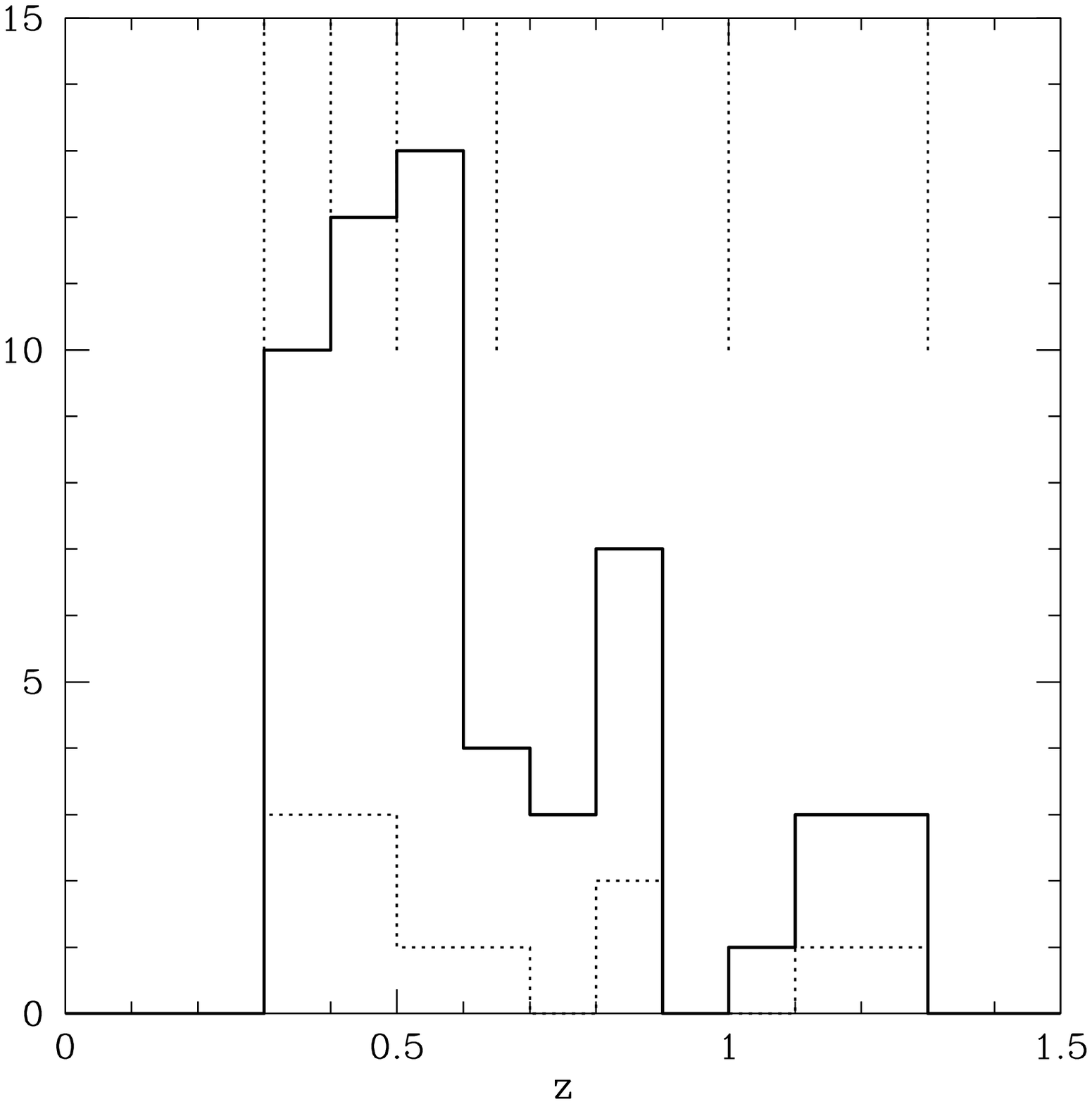}
\caption{Histogram of the redshift distribution of the sample. The
solid line refers to the whole sample, while the dotted line displays
only clusters with $kT\leq5$ keV (see Table~\ref{results}).  The
vertical dotted lines indicate the five redshift intervals selected
for the combined spectral analysis, as described in the text.}
\label{zhist}
\end{figure}

\begin{figure}
\centering
\includegraphics[width=8.0 cm, angle=0]{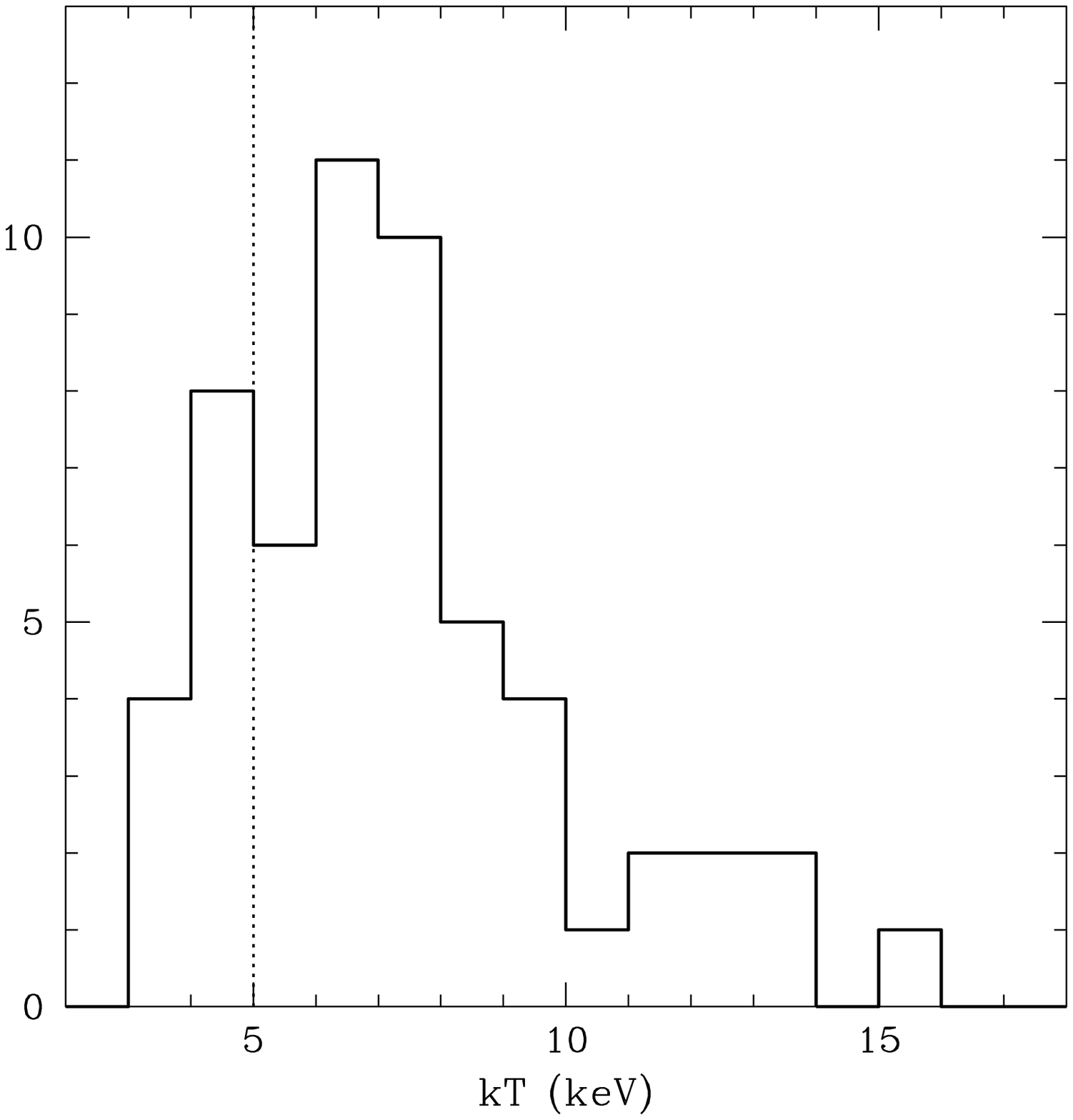}
\caption{Histogram of the temperature distribution of the sample. The
vertical dotted line separates the moderate temperature ($kT<5$ keV)
clusters from the high temperature ($kT>5$ keV) clusters.}
\label{Thist}
\end{figure}

This section presents the results of the spectral analysis of each
of the 56 clusters of our sample. The results of the spectral fits,
referring to the region of radius $R_{ext}$, defined in Sect.~2.1, are
listed in Table~\ref{results}. Histograms of the redshift and
temperature distribution of the sample are shown in Figs.~\ref{zhist}
and \ref{Thist}, respectively.

\begin{table*}
\caption{\label{results}Spectral fit results obtained with the 
{\tt tbabs(mekal)} model. }
\centering
\begin{tabular}{l l l l l l l l}
\hline\hline
Cluster & z & kT [keV]$\mathrm{^a}$ & $Z/Z_\odot$ (And. \& Gre.)$\mathrm{^b}$ & 
$Z/Z_\odot$ (Aspl.)$\mathrm{^c}$ & $N_H$ [cm$^{-2}$]$\mathrm{^d}$ & 
$\chi^2_r$ [d.o.f.] $\mathrm{^e}$ & Null-Hyp. Prob.$\mathrm{^f}$  \\
\hline
MS $1008.1-1224$        &  0.306  & $5.8_{-0.2}^{+0.3}$    & $0.30_{-0.06}^{+0.07}$  & $0.47_{-0.10}^{+0.11}$ & $7.26\times10^{20}$ & $1.23$ [228] &0.009  \\
MS $2137.3-2353$        &  0.313  & $4.96\pm 0.11$	   & $0.35\pm 0.03$	     & $0.56_{-0.05}^{+0.06}$ & $3.55\times10^{20}$ & $1.23$ [289] &0.004  \\
Abell 1995              &  0.319  & $8.60\pm 0.32$	   & $0.40\pm 0.06$	     & $0.64_{-0.10}^{+0.10}$ & $1.42\times10^{20}$ & $1.22$ [343] &0.003  \\
MACS J$0308.9+2645$     &  0.324  & $11.2\pm 0.7$	   & $0.37\pm 0.06$	     & $0.70_{-0.12}^{+0.10}$ & $1.18\times10^{21}$ & $1.07$ [283] &0.194  \\
ZwCl $1358.1+6245$      &  0.328  & $6.70\pm 0.26$	   & $0.40\pm 0.06$	     & $0.64_{-0.10}^{+0.11}$ & $1.92\times10^{20}$ & $1.23$ [282] &0.006  \\ 
MACS J$0404.6+1109$     &  0.355  & $6.9_{-0.8}^{+0.6}$    & $0.16_{-0.11}^{+0.07}$  & $0.22_{-0.20}^{+0.14}$ & $1.43\times10^{21}$ & $1.01$ [165] &0.451  \\
RX J$0027.6+2616$       &  0.367  & $9.1_{-1.5}^{+2.6}$    & $0.57_{-0.19}^{+0.27}$  & $1.02_{-0.35}^{+0.34}$ & $3.86\times10^{20}$ & $1.36$ [52]  &0.042  \\
MACS J$1720.2+3536$     &  0.391  & $6.30 \pm 0.33$	   & $0.50_{-0.06}^{+0.05}$  & $0.83_{-011}^{+0.07}$  & $3.40\times10^{20}$ & $0.74$ [189] &0.997  \\
ZwCl $0024.0+1652$      &  0.395  & $4.38 \pm 0.27$	   & $0.75_{-0.18}^{+0.20}$  & $1.22_{-0.29}^{+0.32}$ & $4.20\times10^{20}$ & $1.02$ [128] &0.410  \\
V $1416+4446$           &  0.400  & $3.50 \pm 0.18$	   & $0.88\pm 0.18$	     & $1.50_{-0.27}^{+0.35}$ & $1.29\times10^{20}$ & $1.11$ [98]  &0.219  \\
MACS J$0159.8-0849$     &  0.405  & $9.2_{-0.5}^{+0.6}$    & $0.36\pm 0.05$	     & $0.59_{-0.09}^{+0.08}$ & $2.08\times10^{20}$ & $1.13$ [208] &0.097  \\
MACS J$2228.5+2036$     &  0.412  & $7.9 \pm 0.6$          & $0.41_{-0.07}^{+0.06}$  & $0.59_{-0.10}^{+0.09}$ & $4.58\times10^{20}$ & $0.91$ [196] &0.801  \\
MS $0302.7+1658$        &  0.424  & $4.34_{-0.44}^{+0.56}$ & $0.40_{-0.18}^{+0.21}$  & $0.58_{-0.28}^{+0.34}$ & $1.11\times10^{21}$ & $0.98$ [31]  &0.490  \\
MS $1621.5+2640$        &  0.426  & $6.9_{-0.6}^{+0.7}$    & $0.35_{-0.09}^{+0.08}$  & $0.63 \pm 0.14$        & $3.58\times10^{20}$ & $0.90$ [157] &0.804  \\
MACS J$0417.5-1154$     &  0.440  & $11.3\pm 0.8$	   & $0.29_{-0.07}^{+0.06}$  & $0.45_{-0.11}^{+0.10}$ & $3.86\times10^{20}$ & $0.82$ [209] &0.971  \\
MACS J$1206.2-0847$     &  0.440  & $11.0_{-0.6}^{+0.7}$   & $0.18_{-0.06}^{+0.05}$  & $0.25_{-0.09}^{+0.10}$ & $3.72\times10^{20}$ & $1.15$ [271] &0.045  \\
RX J$1347.5-1145$       &  0.451  & $14.0\pm 0.4$	   & $0.33\pm 0.04$	     & $0.55_{-0.06}^{+0.07}$ & $4.92\times10^{20}$ & $1.20$ [436] &0.003  \\
V $1701+6414$           &  0.453  & $4.27_{-0.25}^{+0.26}$ & $0.54_{-0.12}^{+0.13}$  & $0.85_{-0.20}^{+0.22}$ & $2.46\times10^{20}$ & $0.95$ [123] &0.649  \\
CL $1641+4001$          &  0.464  & $4.8 \pm 0.6$          & $0.48_{-0.16}^{+0.19}$  & $0.79_{-0.24}^{+0.29}$ & $1.10\times10^{20}$ & $0.98$ [53]  &0.509  \\
MACS J$1621.4+3810$     &  0.465  & $6.5 \pm 0.7$ 	   & $0.13 \pm 0.08$	     & $0.19_{-0.16}^{+0.15}$ & $1.09\times10^{20}$ & $1.46$ [67]  &0.549  \\
MACS J$1824.3+4309$     &  0.487  & $7.2_{-1.3}^{+2.2}$    & $0.38_{-0.25}^{+0.23}$  & $ 0.55_{-0.33}^{+0.31}$& $4.46\times10^{20}$ & $1.03$ [33]  &0.423  \\
MACS J$1311.0-0311$     &  0.492  & $8.0 \pm 0.9$ 	   & $0.39 \pm 0.09$	     & $0.60_{-0.19}^{+0.21}$ & $1.87\times10^{20}$ & $0.97$ [87]  &0.571  \\
V $1525+0958$           &  0.516  & $5.4_{-0.5}^{+0.4}$    & $0.35 \pm 0.10$	     & $0.54_{-0.20}^{+0.21}$ & $2.91\times10^{20}$ & $1.06$ [107] &0.315  \\
MS $0451.6-0305$        &  0.539  & $8.2_{-0.3}^{+0.4}$    & $0.34\pm0.06$	     & $ 0.57 \pm 0.10$       & $4.97\times10^{20}$ & $1.24$ [279] &0.004  \\
MS $0015.9+1609$        &  0.541  & $9.3_{-0.3}^{+0.5}$    & $0.33_{-0.05}^{+0.06}$  & $ 0.53 \pm 0.09$       & $4.07\times10^{20}$ & $1.01$ [302] &0.460  \\
MACS J$1149.5+2223$     &  0.544  & $12.9_{-1.0}^{+1.2}$   & $0.21_{-0.07}^{+0.06}$  & $0.37_{-0.11}^{+0.10}$ & $2.28\times10^{20}$ & $1.04$ [254] &0.317  \\
MACS J$1423.8+2404$     &  0.545  & $7.3_{-0.5}^{+0.6}$    & $0.31_{-0.08}^{+0.06}$  & $0.46_{-0.10}^{+0.13}$ & $2.38\times10^{20}$ & $1.11$ [135] &0.174  \\
MACS J$0717.5+3745$     &  0.548  & $13.3 \pm 0.7$         & $0.18_{-0.04}^{+0.05}$   & $0.29_{-0.07}^{+0.11}$ & $7.04\times10^{20}$ & $1.14$ [404] &0.029  \\
V $1121+2327$           &  0.562  & $5.2\pm 0.5$           & $0.27_{-0.08}^{+0.10}$  & $0.43_{-0.07}^{+0.15}$ & $1.32\times10^{20}$ & $0.99$ [102] &0.499  \\
SC $1120-1202$          &  0.562  & $5.7_{-0.8}^{+1.3}$    & $0.23_{-0.17}^{+0.20}$  & $0.33_{-0.21}^{+0.33}$ & $5.19\times10^{20}$ & $1.13$ [45]  &0.256  \\
RX J$0848.7+4456$       &  0.570  & $3.4 \pm 0.30$	   & $0.51_{-0.19}^{+0.23}$  & $0.79_{-0.30}^{+0.36}$ & $2.63\times10^{20}$ & $1.33$ [46]  &0.068  \\
MACS J$2129.4-0741$     &  0.570  & $8.7_{-0.8}^{+0.7}$    & $0.51_{-0.11}^{+0.08}$  & $0.82 \pm 0.14$        & $4.82\times10^{20}$ & $0.85$ [132] &0.890  \\
MS $2053.7-0449$        &  0.583  & $5.68_{-0.47}^{+0.57}$ & $0.16_{-0.10}^{+0.11}$  & $0.24_{-0.16}^{+0.18}$ & $5.02\times10^{20}$ & $1.03$ [100] &0.402  \\
MACS J$0647.7+7015$     &  0.584  & $15.5_{-1.7}^{+2.3}$   & $<0.10$		     & $< 0.15$ 	      & $5.64\times10^{20}$ & $0.91$ [118] &0.737  \\
RX J$0956.0+4107$       &  0.587  & $7.4^{2.5}_{-1.45}$    & $0.13_{-0.13}^{+0.28}$  & $0.21_{-0.21}^{+0.45}$ & $1.14\times10^{20}$ & $1.08$ [27]  &0.353  \\
CL $0542.8-4100$        &  0.634  & $7.9_{-0.8}^{+1.1}$    & $0.20_{-0.09}^{+0.12}$ & $0.31_{-0.15}^{+0.19}$ & $3.73\times10^{20}$ & $1.19$ [113] &0.081  \\
RCS J$1419.2+5326$      &  0.640  & $4.1_{-0.6}^{+0.7}$    & $0.15_{-0.15}^{+0.21}$  & $0.16_{-0.16}^{+0.31}$ & $1.18\times10^{20}$ & $1.35$ [21]  &0.131  \\
MACS J$0744.9+3927$     &  0.686  & $9.2 \pm 0.6$          & $0.28 \pm 0.06$	     & $0.46_{-0.12}^{+0.10}$ & $5.71\times10^{20}$ & $1.22$ [193] &0.021  \\
RX J$1221.4+4918$       &  0.700  & $8.4_{-0.8}^{+0.9}$    & $0.29_{-0.12}^{+0.13}$  & $0.48_{-0.19}^{+0.21}$ & $1.47\times10^{20}$ & $0.99$ [141] &0.534  \\
RX J$1113.1-2615$       &  0.730  & $5.7_{-0.6}^{+0.9}$    & $0.43_{-0.17}^{+0.20}$  & $0.68_{-0.27}^{+0.30}$ & $5.50\times10^{20}$ & $0.69$ [59]  &0.966  \\
RX J$2302.8+0844$       &  0.734  & $8.0_{-1.1}^{+1.3}$    & $0.12_{-0.12}^{+0.16}$  & $0.19_{-0.19}^{0.26}$  & $4.85\times10^{20}$ & $0.85$ [81]  &0.833  \\
MS $1137.5+6624$        &  0.782  & $6.8 \pm 0.5$	   & $0.26 \pm 0.11$	     & $0.43_{-0.17}^{+0.18}$ & $1.21\times10^{20}$ & $1.09$ [151] &0.204  \\
RX J$1317.4+2911$       &  0.805  & $4.5_{-1.0}^{+1.4}$    & $0.35_{-0.30}^{+0.46}$  & $0.49_{-0.46}^{+0.67}$ & $1.10\times10^{20}$ & $0.71$ [14]  &0.768  \\
RX J$1350.0+6007$       &  0.810  & $4.4_{-0.6}^{+0.7}$    & $0.56_{-0.18}^{+0.27}$  & $0.96_{-0.33}^{+0.35}$ & $1.80\times10^{20}$ & $1.26$ [51]  &0.103  \\
RX J$1716.4+6708$       &  0.813  & $6.9_{-0.7}^{+0.8}$    & $0.49_{-0.16}^{+0.18}$  & $0.79_{-0.26}^{+0.29}$ & $3.72\times10^{20}$ & $0.77$ [76]  &0.934  \\
RX J$0152.7-1357$ S     &  0.828  & $8.7_{-1.8}^{+2.4}$    & $<0.22$		     & $<0.37$  	      & $1.54\times10^{20}$ & $1.06$ [32]  &0.376  \\
MS $1054.4-0321$        &  0.832  & $7.5_{-0.4}^{+0.7}$    & $0.23_{-0.08}^{+0.07}$  & $0.38_{-0.12}^{+0.12}$ & $3.61\times10^{20}$ & $1.15$ [238] &0.056  \\
RX J$0152.7-1357$ N     &  0.835  & $6.7_{-1.0}^{+1.2}$    & $0.17_{-0.16}^{+0.19}$  & $0.25_{-0.25}^{+0.30}$ & $1.54\times10^{20}$ & $0.87$ [45]  &0.715  \\
1WGA J$1226.9+3332$     &  0.890  & $12.9_{-1.2}^{1.4}$	   & $0.02_{-0.02}^{+0.12}$  & $<0.21$  	      & $1.38\times10^{20}$ & $1.06$ [97]  &0.335  \\
CL $1415.1+3612$        &  1.030  & $7.0_{-0.7}^{+0.8}$    & $0.24_{-0.13}^{+0.15}$  & $0.40_{-0.21}^{+0.23}$ & $1.09\times10^{20}$ & $0.79$ [63]  &0.891  \\
RDCS J$0910+5422$       &  1.106  & $6.4_{-1.2}^{+1.5}$    & $<0.14$                 & $<0.21$  	      & $2.10\times10^{20}$ & $0.94$ [28]  &0.559  \\
RX J$1053.7+5735$ E     &  1.134  & $3.4_{-0.4}^{+0.5}$    & $0.51_{-0.15}^{+0.14}$  & $0.58_{-0.22}^{+0.75}$ & $5.7\times10^{19}$  & $1.11$ [29]  &0.317  \\
RX J$1053.7+5735$ W     &  1.134  & $7.2_{-0.9}^{+1.2}$    & $0.32_{-0.15}^{+0.14}$  & $0.51_{-0.24}^{+0.23}$ & $5.7\times10^{19}$  & $0.55$ [31]  &0.981  \\
RDCS J$1252-2927$       &  1.235  & $7.2_{-0.6}^{+0.4}$    & $0.35_{-0.09}^{+0.06}$  & $0.58_{-0.15}^{+0.11}$ & $5.95\times10^{20}$ & $1.01$ [55]  &0.454  \\
RDCS J$0849+4452$       &  1.261  & $6.2_{-0.9}^{+1.0}$    & $0.16_{-0.14}^{+0.13}$  & $0.21_{-0.19}^{+0.27}$ & $2.63\times10^{20}$ & $0.51$ [24]  &0.977  \\
RDCS J$0848+4453$       &  1.273  & $3.4_{-1.1}^{+2.5}$    & $0.22_{-0.22}^{+1.33}$  & $0.22_{-0.22}^{+1.38}$ & $2.63\times10^{20}$ & $1.04$ [11]  &0.411  \\
\hline
\end{tabular}
\begin{list}{}{}
\item[Notes:] $\mathrm{^a}$ temperature; $\mathrm{^b}$ iron abundance in solar units 
by \citet{and89} and $\mathrm{^c}$ by \citet{asp05}; $\mathrm{^d}$ local column 
density, always fixed to the Galactic value by Dickey \& Lockman (1990); 
$\mathrm{^e}$ reduced chi-square and degrees of freedom obtained
after binning the spectra to 20 counts per bin; $\mathrm{^f}$ null-hypothesis 
probability. Errors refer to the $1\sigma$ confidence level. 
\end{list}
\end{table*}

The redshift distribution is peaked around $z\simeq0.5$, while at
$z>1$ we have only 7 objects (we recall that we consider
the two clumps of RX~J1053 separately, see \citealt{has04}). In order to
investigate the properties of the sample as a function of redshift, we
divide the clusters in 5 redshift intervals (10 objects with
$0.3<z<0.4$; 12 objects with $0.4<z<0.5$; 15 objects with
$0.5<z<0.65$; 12 objects with $0.65<z<1.0$, and 7 objects with $1.0
<z< 1.3$). These intervals (Fig.~\ref{zhist}) are defined in order 
to have a comparable number of objects in a reasonably narrow redshift range.

\begin{figure}
\centering
\includegraphics[width=8.0 cm, angle=0]{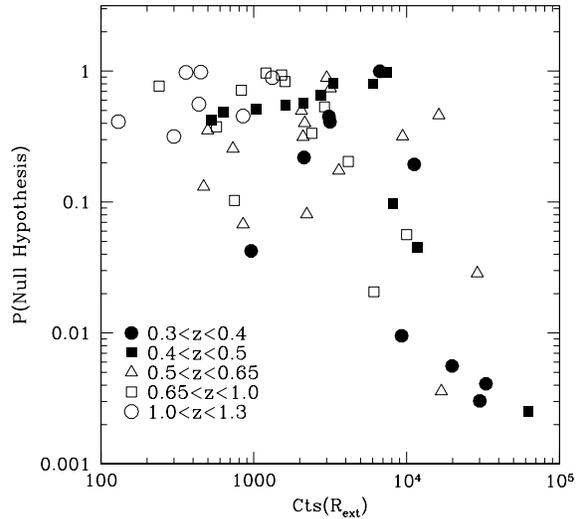}
\caption{Null-hypothesis probability for the single-temperature
best fits as a function of net detected counts for the whole sample.}
\label{nhyp}
\end{figure}

As shown in the temperature distribution (Fig.~\ref{Thist}), we sampled
mostly hot clusters ($kT > 5$ keV), while only 12 are in the {\sl
medium} temperature range ($3 < kT < 5$ keV). We would like to point out 
here that we derived a single spectral temperature for the region within
$R_{ext}$. In principle, the spectral temperature can be significantly
different from the emission-weighted and gas-mass-weighted
temperature, in the presence of a thermally complex ICM
\citep[e.g.][]{maz04}. In a few cases, the effect of the temperature
gradient is strong enough to make the single-temperature fit
unacceptable.  To evaluate the goodness of each fit, we computed the
$\chi^2$ of each best-fit model after binning the spectrum to 20
counts per bin. We find that most of our spectra are well-fitted by a
single-temperature {\tt mekal} model (see 7th column in
Table~\ref{results}).  However, when the number of the net detected
counts becomes larger than $\sim 10^4$, the quality of the fits drops
dramatically for about half of the clusters (see Fig.~\ref{nhyp}).  If
we consider a 1\% null-hypothesis probability as the threshold for an
acceptable fit, then we must reject the single-temperature model
within $R_{ext}$ for 5 clusters in our sample (namely MS~2137, A~1995,
ZW~1358, RX~J1347, and MS~0451). Note that we can also fit clusters 
with very disturbed morphology and high S/N (e.g. MACS~J0717) with a
single-temperature model, due to the very high temperatures
involved, which provide composite spectra with much less features than
spectra with low-temperature components. Indeed, the
single-temperature model fails mostly when a strong cool-core with
temperatures lower than 3 keV is present \citep[e.g.][]{maz04}, while
it is still acceptable if the temperature range is well above this
threshold.

Since we are focusing here on metallicity, we made a closer
investigation of the best-fit $Z_{Fe}$ values for the clusters with
the lowest null-hypothesis probability. Since these clusters are also
the ones with the highest S/N, we were able to perform a spatially
resolved spectral analysis for about four concentric annuli.  We find
that the best-fit $Z_{Fe}$ value measured with a single temperature
mekal model within $R_{ext}$, is representative of the inner 400~kpc
and is not dominated by the central bin.  This result, implying that
the presence of temperature gradient does not dramatically affect the
measurements of iron abundance, is reinforced by the spectral
simulations described in the Appendix.  Therefore we used the
single-temperature best-fit values for all the clusters in our
sample. The attempt to model the evolution of $Z_{Fe}$ separately in
the inner 100~kpc and the outer regions is deferred to a future paper.

\begin{figure}
\centering
\includegraphics[width=8.0 cm, angle=0]{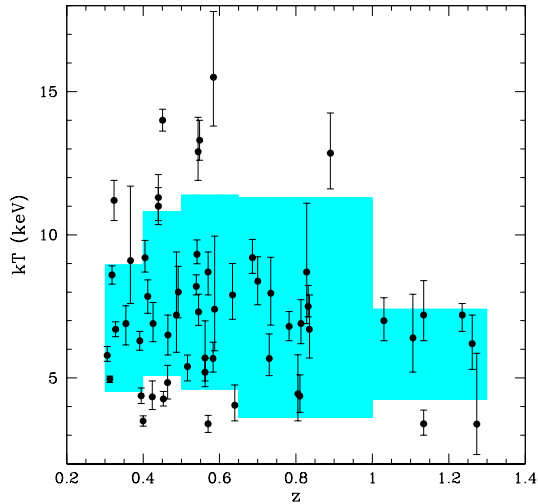}
\caption{Temperature plotted vs redshift for the whole sample. Shaded
areas show the {\sl rms} dispersion around the weighted mean.}
\label{tempz}
\end{figure}

Finally, we show in Fig.~\ref{tempz} the distribution of
temperatures in our sample as a function of redshifts (error bars are
at the $1\sigma$ confidence level). The Spearman test shows no correlation
between temperature and redshift (Spearman's rank coefficient of $r_s
=-0.095$ for 54 degrees of freedom, probability of null correlation
$p=0.48$).  Fig.~\ref{tempz} shows that the range of temperatures in 
each redshift bin is about $6-7$~keV. 
Therefore, we are sampling a population of medium-hot clusters uniformly 
with redshift, with the hottest clusters preferentially in the redshift 
bin $0.4<z<0.6$.

\begin{figure}
\centering
\includegraphics[width=5.4 cm, angle=0]{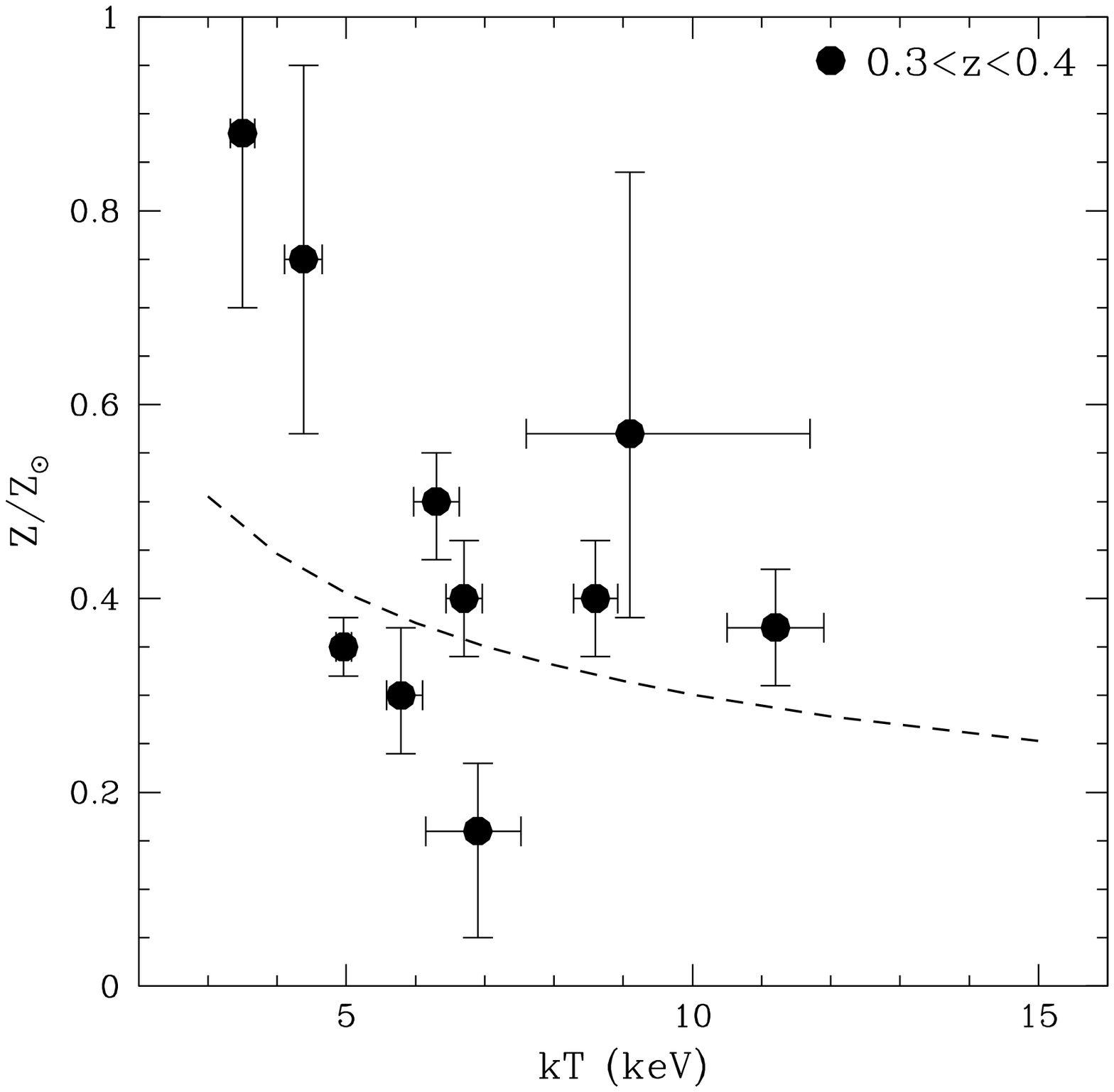}
\includegraphics[width=5.4 cm, angle=0]{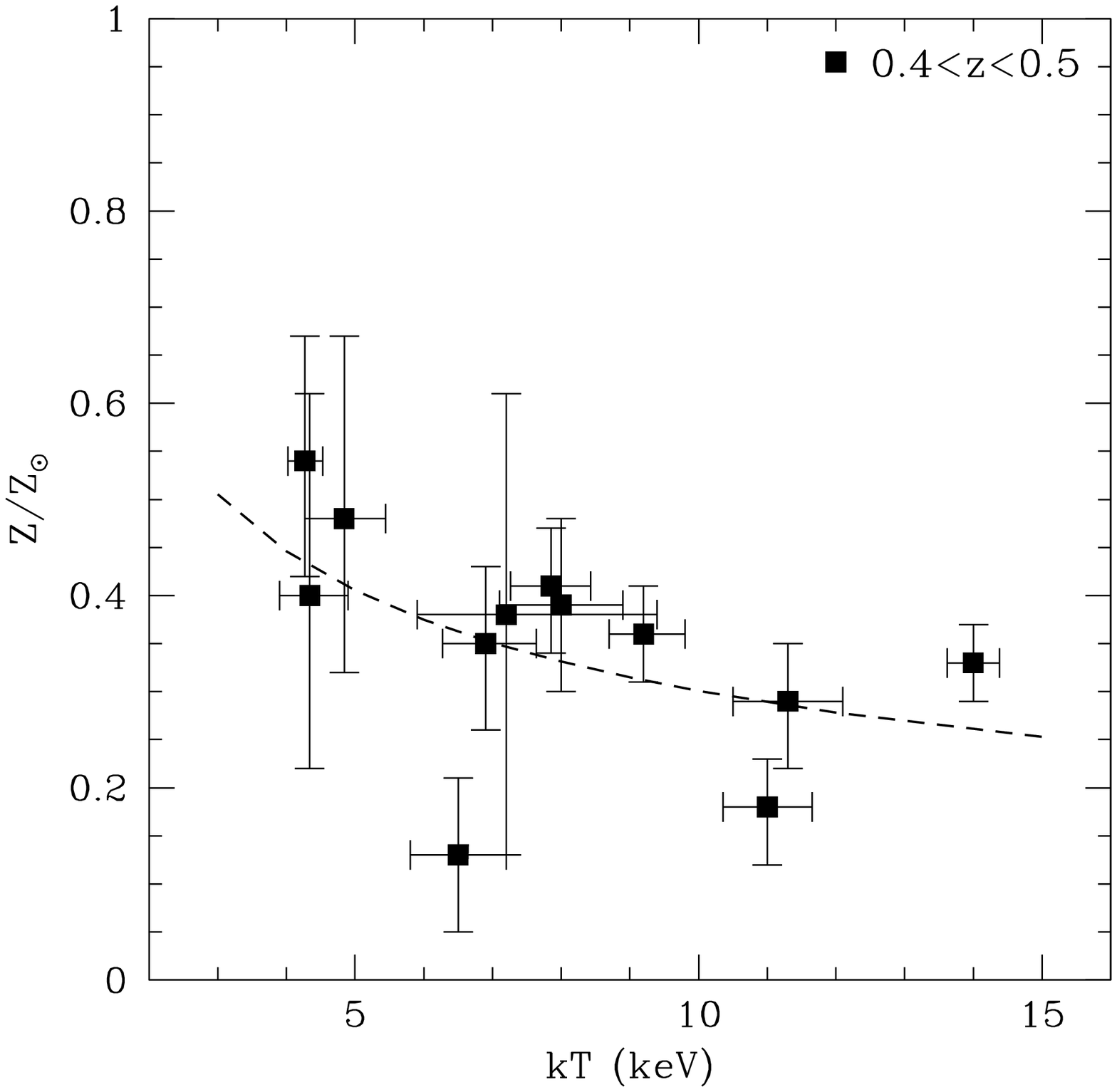}
\includegraphics[width=5.4 cm, angle=0]{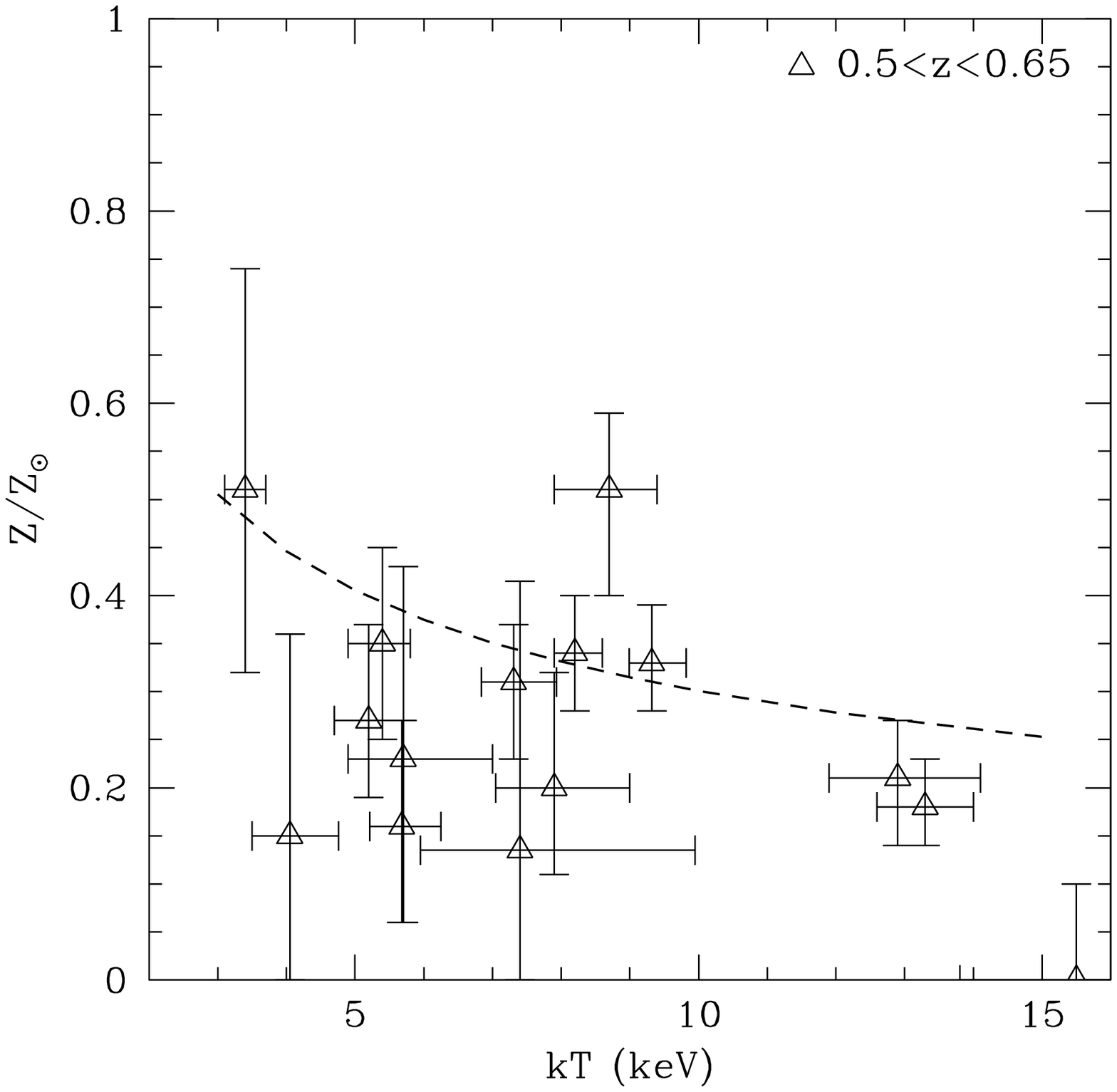}
\includegraphics[width=5.4 cm, angle=0]{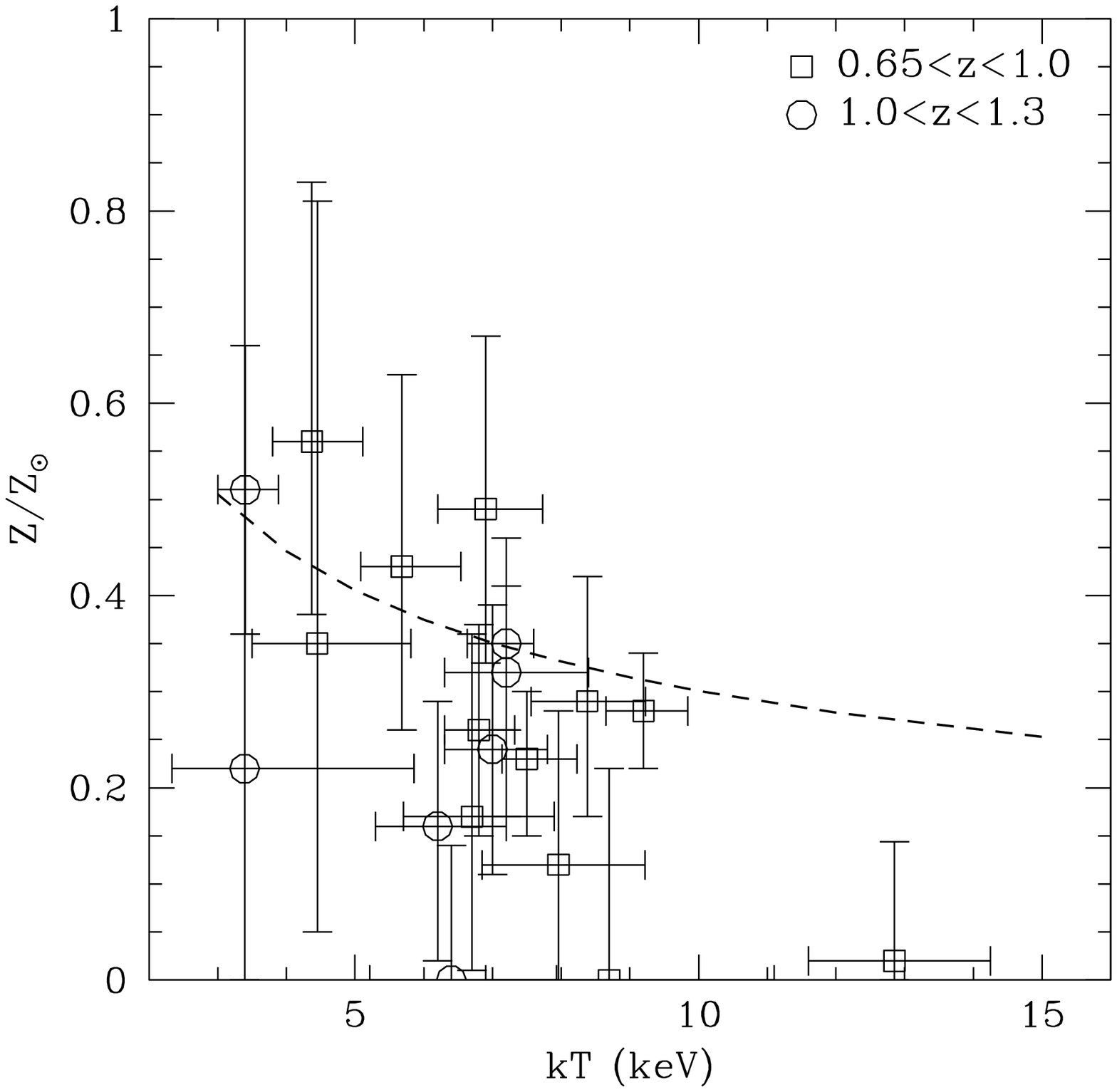}
\caption{Iron abundance-temperature plots for the whole sample. The
four panels show each redshift bin separately (last two bins in the
fourth panel). The dashed line represents the best-fit
metallicity-temperature relation ($Z/Z_\odot\simeq0.88 \,T^{-0.47}$)
referring to the whole sample. Error bars refer to $1\sigma$
confidence level.}
\label{tmbins}
\end{figure}

The relations between temperature and iron abundance at different
redshifts are shown in Fig.~\ref{tmbins}.  For three clusters we can
only derive upper limits on the iron abundance of the ICM, two of them
at $z<0.8$. For five clusters we measure positive iron abundances, 
which are still consistent with no detection at the $1\sigma$ c.l..
Overall, we detect the presence of the iron line in the large majority
of the clusters, and measure the iron abundance with a typical error
of 30\% at the $1\sigma$ c.l. for $z<0.6$, and 50\% or larger at $z>0.6$.

\subsection{The iron abundance-temperature correlation}

\begin{figure}
\centering
\includegraphics[width=8.0 cm, angle=0]{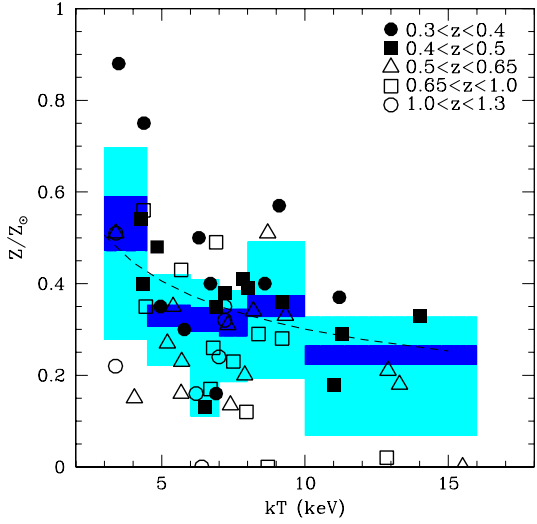}
\caption{Scatter plot of best-fit iron abundance values (without error
bars) versus temperature for the whole sample. The dashed line
represents the best-fit metallicity-temperature relation
($Z/Z_\odot\simeq0.88\,T^{-0.47}$).  Shaded areas show the weighted
mean (blue) and average iron abundance with {\em rms} dispersion
(cyan) in 6 temperature bins (see Table~\ref{tableT}).}
\label{tm_all}
\end{figure}

Our analysis (Fig.~\ref{tm_all}) suggests higher iron abundances at
lower temperatures in all the redshift bins. This trend is somewhat
blurred by the large scatter. We find a more than $2\sigma$ negative
correlation for the whole sample, with a Spearman's rank coefficient
of $r_s =-0.31$ for 54 degrees of freedom (probability of no
correlation $p=0.018$).  The correlation is more evident when we
compute the weighted average of the metallicity in six temperature
intervals (see Table~\ref{tableT}), as shown by the shaded areas in
Fig.~\ref{tm_all}.  The weighted mean is computed as $T_{wa} =\,
\Sigma T_i w_i$ in each temperature bin, where $w_i =
1/\sigma_i^2/\Sigma (1/\sigma_i^2)$, and $\sigma_i$ is the $1\sigma$
error on the single measurement.  We find that in each bin the scatter
of the best-fit values around the mean is comparable to the
statistical errors on the single measurements (reduced
$\chi_\nu^2\simeq 1$ assuming a constant $Z_{Fe}$ in the bin), with
the exception of the third and sixth bins, where the intrinsic scatter
is larger ($\chi_\nu^2\simeq3$).

\begin{table}
\caption{Average iron abundance calculated in different temperature bins.}
\label{tableT}
\centering
\begin{tabular}{l l l}
\hline\hline
$kT$ [keV]$\mathrm{^a}$ & $Z/Z_\odot\mathrm{^b}$ & $\Delta Z/Z_\odot\mathrm{^c}$ \\
 & (weighted mean) & ({\em rms}) \\
\hline

$3.9$ [10]  & $0.542\pm 0.060$ & $0.23$ \\
$5.4$ [8]   & $0.329\pm 0.024$ & $0.10$ \\
$6.7$ [11]    & $0.322\pm 0.027$ & $0.17$ \\
$7.5$ [10]    & $0.317\pm0.030$  & $0.11$ \\
$8.8$ [9]    & $0.350\pm0.024$  & $0.16$ \\
$12.7$ [8]   & $0.244\pm0.021$  & $0.14$ \\ 
\hline
\end{tabular}
\begin{list}{}{}
\item[Notes:] $\mathrm{^a}$ average temperature in each bin (the number of
clusters in each bin is shown in parenthesis); $\mathrm{^b}$ weighted mean 
of the iron abundance; $\mathrm{^c}$ {\em rms} dispersion. 
\end{list}
\end{table}

\begin{figure}
\centering
\includegraphics[width=8.0 cm, angle=0]{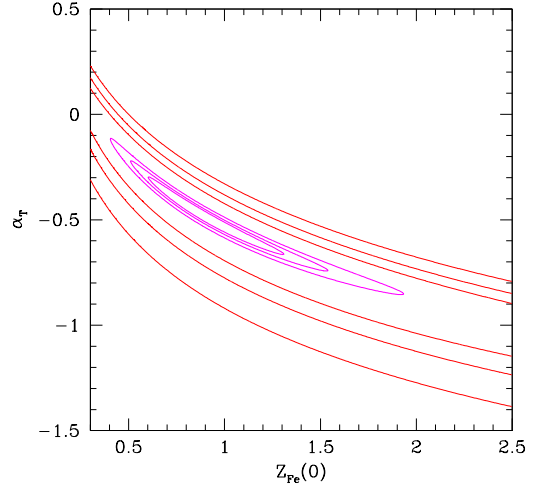}
\caption{Best-fit confidence contour plot for the $Z-T$ relation
modelled with a power law of the form $Z(T)=Z_{Fe}(0)\,
T^{-\alpha_T}$. Inner contours display the 1, 2, and $3\sigma$ c.l. using
errors on the weighted mean, while outer thick contours are obtained
using the {\em rms} dispersion.}
\label{cont_tempmet} 
\end{figure}

This trend is similar to what is found in the ASCA data of nearby
clusters by \citealt{bag05} \citep[see also][]{arn92, mus97,
fin01}. In their paper, the average iron abundance for $kT > 5$ keV is
constant and equal to $Z\simeq 0.3\,Z_\odot$, while it rises to very
high values (well above $0.4\,Z_\odot$) in the temperature range
$2-3$~keV, and drops below $0.3\,Z_\odot$ for $kT< 2$~keV. It is worth
noting, however, that this behavior is different from that of Ni and
$\alpha$-elements. Here, we confirm that in our high-redshift sample
the average measured iron abundance rises below 5~keV.  A simple
power law fit of the form

\begin{equation}
Z(T)=Z_{Fe}(0)\,T^{-\alpha_T}
\end{equation}
\noindent
to our data gives $Z_{Fe}(0)= 0.88^{+0.5}_{-0.2}\,Z_\odot$ and
$\alpha_T = -0.47^{+0.1}_{-0.2}$ with $\chi_\nu^2\simeq1.8$ for 5
degrees of freedom (see the confidence contours in the
slope-normalization plane in Fig.~\ref{cont_tempmet}). This trend is
consistent with what is found from ASCA data of a local sample of
clusters for $kT > 3$~keV \citep[see][]{bag05,hor05}.
Interestingly, a similar correlation between temperature and
metallicity is observed when a spatially-resolved analysis of the ICM
can be performed in a single object, as in the case of the core of the
Perseus cluster \citep[see][]{san04}. However a physical explanation
for this behaviour is still missing.

\begin{figure}
\centering
\includegraphics[width=6.5 cm, angle=0]{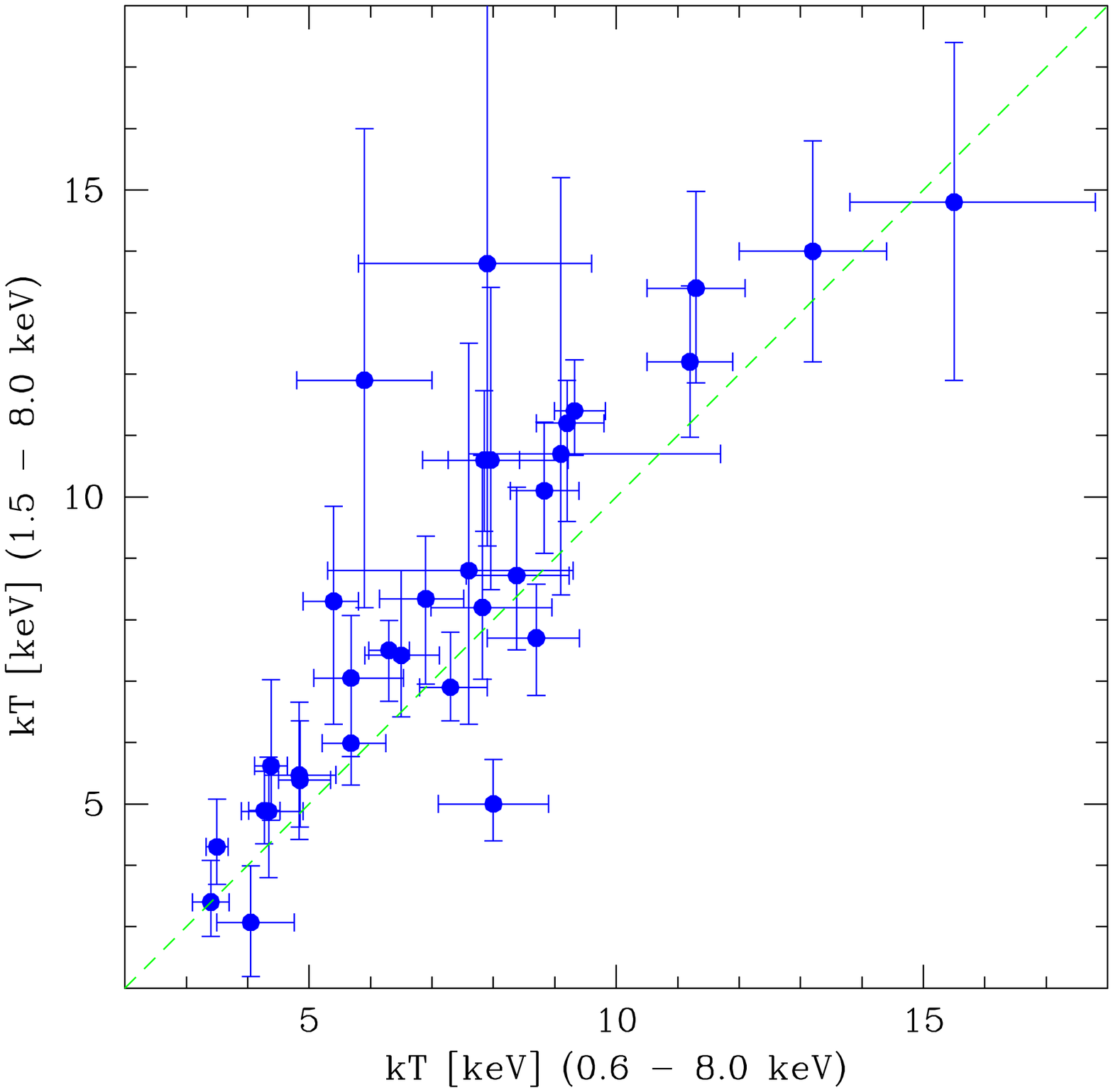}
\includegraphics[width=6.5 cm, angle=0]{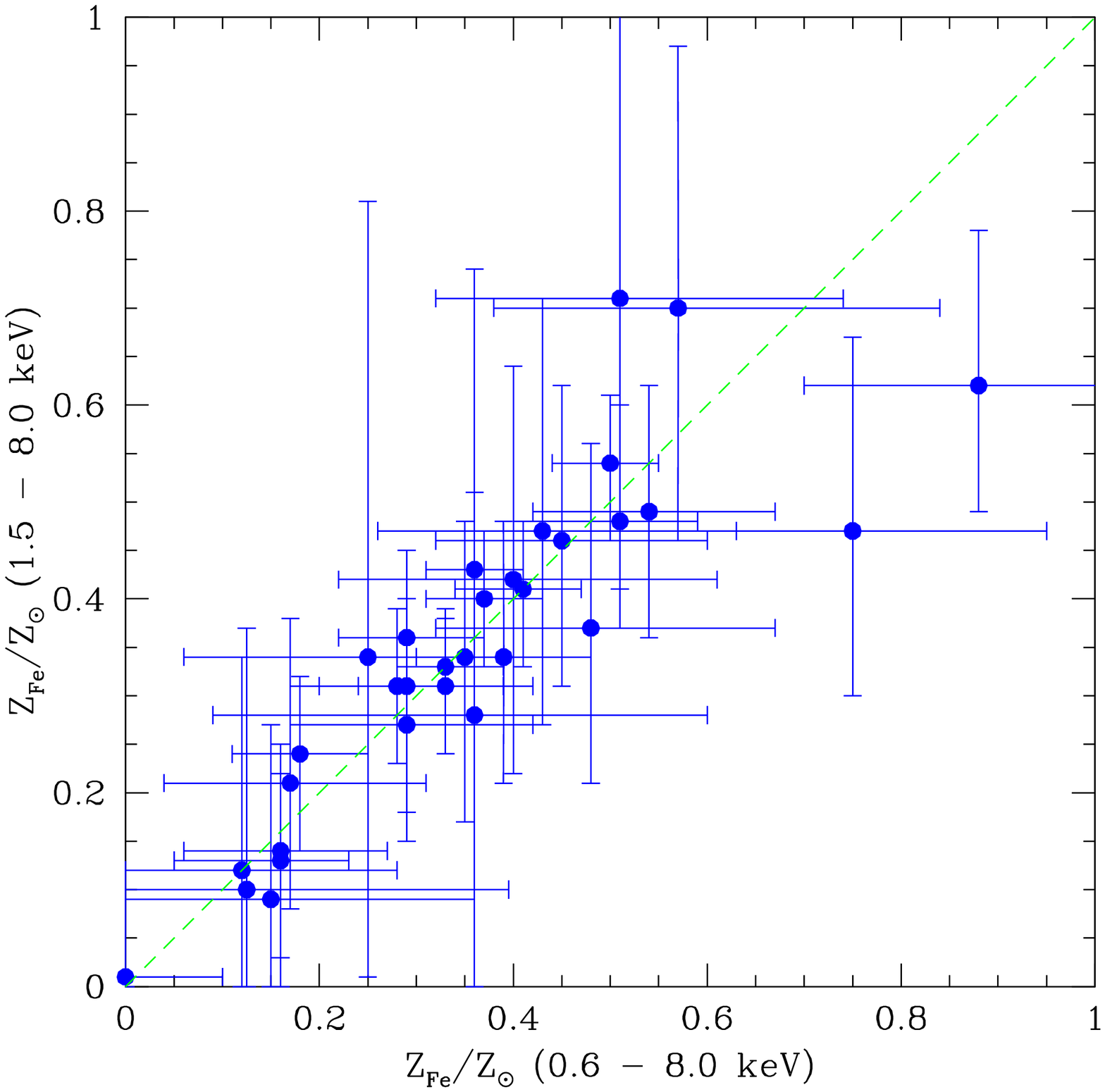}
\caption{Best-fit temperatures ({\em upper panel}) and iron abundances 
({\em lower panel}) obtained using the $1.5-8$~keV energy range compared 
to the values obtained from the $0.6-8$~keV (this paper). 
Dashed lines show the locus of equal 
temperatures and abundances. Only clusters with more than 2000 net counts 
(see Table~\ref{exposures}) are considered here.}
\label{highE}
\end{figure}

We recall that the measure of the iron abundance in local clusters is
based on both the K-shell and the L-shell complex, at $6.7-6.9$~keV
and between 1 and 2~keV, respectively. It has been pointed out that a
diagnostic based mostly on the L-shell, as in the case of spectra with
a significant low temperature component, is more uncertain
\citep[]{ren97}.  In our high redshift sample, we expect to be
sensitive only to the iron K-shell complex. Indeed, when we fit the
spectra cutting energies below 2~keV rest-frame, we find that the
best-fit metallicities are consistent with those found within the
$0.6-8$~keV range (observed frame) used throughout the paper, as shown
in Fig.~\ref{highE}. We notice that when the low energy range is
removed, errors on the temperatures become larger, with a clear
tendency to higher values. This trend is expected since higher
temperature spectral shapes can be accommodated more easily than lower
ones.  On the other hand, iron abundances hardly change, except for
the two most iron-rich clusters, ZW~0024 and V~1416. This may indicate
that in these two objects the high iron abundance could be associated
with a low temperature component. However, in these cases a separate
analysis in two annuli is possible and we do not find a clear
enhancement of the iron abundance in the inner regions.

The observed trend between $Z_{Fe}$ and $T$ is still a matter of debate. 
It may be linked to the observed decrease in the star formation efficiency 
with increasing cluster mass as reported in \citet[]{lin03}. 
The modelization of this trend goes beyond the scope of this paper.

\subsection{The evolution of the iron abundance via combined spectral 
analysis}

\begin{figure}
\centering
\includegraphics[width=8.0 cm, angle=0]{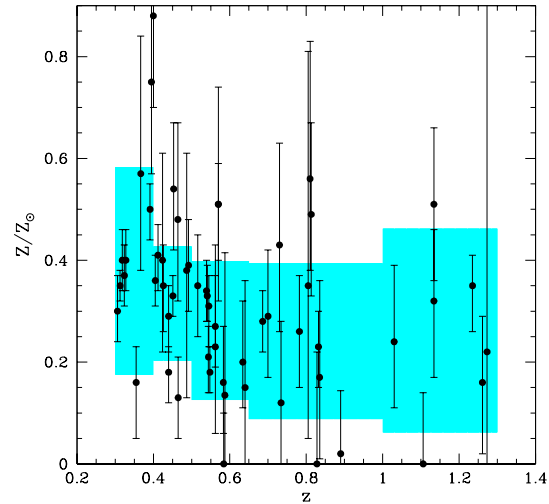}
\caption{Iron abundance plotted versus redshift for the 56 clusters of
the sample. Shaded areas show the {\em rms} dispersion around the
weighted mean of the iron abundance in the 5 redshift bins defined in
the text.  Error bars refer to a $1\sigma$ confidence level.}
\label{metz}
\end{figure}

In Fig.~\ref{metz} we show the iron abundance best-fit values for
all the sources in the sample. When focusing on the highest redshift bin, it
is worth noticing that at $z>1$ we find clear detections of the Fe K
line in the spectra of CL~J1415 ($z=1.030$, $Z_{Fe} >0$ at the 90\% c.l.),
of RDCS~J1252 ($z=1.235$, $Z_{Fe} >0.2 \, Z_\odot$ at the 90\% c.l.), and of the
two clumps of RX~J1053 ($z=1.134$, $Z_{Fe} >0.1 \, Z_\odot$ at the 90\%
c.l.). In the spectra of the four other clusters, we do not have
separate detections of the iron line, but all measurements are
consistent within $1\sigma$ with $Z_{Fe}\simeq0.3\,Z_\odot$. Therefore, 
at present, we have a much better estimate
of the metal content of clusters at $z \simeq 1$ than in
Paper~I, where the iron line at $z>1$ was only firmly detected in the
two clumps of RX~J1053.

We find a $\sim 3\sigma$ negative correlation between iron abundance
and redshift, with a Spearman's rank coefficient of $r_s =-0.40$ for
54 degrees of freedom (probability of a null correlation $p=0.0023$).
This correlation is stronger than the weak hint (less than a $2\sigma$
c.l.) of anticorrelation found in Paper~I. We verified that, if we
calculate the Spearman's rank coefficient only for the 19 clusters
analyzed in Paper~I, using the newly revised temperature and
abundances (consistent with the old ones as shown in
Fig.~\ref{compare}), we again obtain very weak correlation ($r_s
=-0.25$ for 17 degrees of freedom, probability of no correlation
$p=0.30$), which is therefore consistent with the results reported in
Paper~I. This confirms that our new results should not be ascribed to
updated calibrations, but rather to the larger size of the sample,
particularly at $z<0.5$.

The decrease in $Z_{Fe}$ with redshift becomes more evident by
computing the average iron abundance as determined by a {\em combined}
spectral fit in a given redshift bin.  This technique is similar to
the stacking analysis often performed in optical spectroscopy, where
spectra from a homogeneous class of sources are averaged together to
boost the S/N, thus allowing the study of otherwise undetected
features. In our case, different X-ray spectra cannot be stacked due
to their different shape (different temperatures).  Therefore, a
simultaneous spectral fit is performed leaving temperature and
normalization free to vary for each object, and using a unique
metallicity value for all the clusters in the subsample.

We note that the scatter of the best-fit values around the mean in
each redshift bin is, in some cases, larger than the typical
statistical errors on single measurements. This is expected on the
basis of the $Z-T$ correlation, as found in Sect.~3.3.  The reduced
$\chi^2$ obtained by assuming that measurements are scattered around the
weighted average is between 2 and 3 in the first two bins, implying
the presence of intrinsic scatter comparable to the typical
statistical error, while it is $\sim1$ above $z\simeq0.5$, since here
the typical statistical error is larger than the intrinsic scatter
component.  By assuming a unique value of $Z_{Fe}$ in the combined
fit, however, we intend to provide an average value of the metallicity
over large cluster volumes as a function of redshift.

\begin{figure}
\centering
\includegraphics[width=8.0 cm, angle=0]{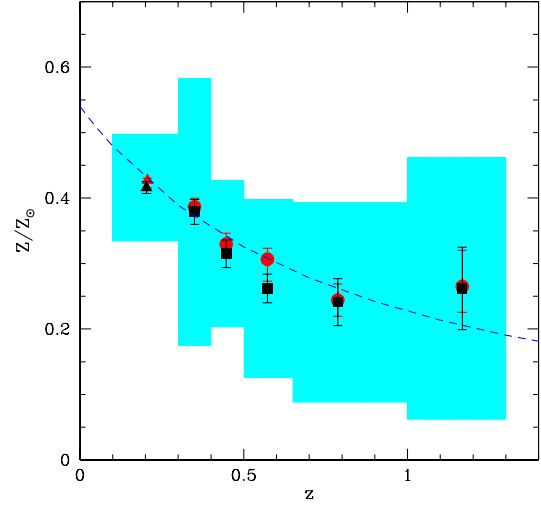}
\caption{Mean iron abundance from combined fits within five redshift
bins defined in the text (red circles) compared with the weighted
average of single-source measurements in the same bins (black
squares). The triangles at $z\simeq0.2$ are based on the low-z sample
described in Sect.~3.5.  Error bars refer to the $1\sigma$ confidence level.
Shaded areas show the {\em rms} dispersion.  The dashed line indicates
the best fit over the 6 redshift bins for a simple power law of the
form $\langle Z_{Fe}\rangle=Z_{Fe}(0)\,(1+z)^{-\alpha_z}$ with
$\alpha_z\simeq 1.25$.}
\label{Z_vs_z} 
\end{figure}

A plot of the combined iron abundance measured in each redshift bin is
shown in Fig.~\ref{Z_vs_z}. To verify the robustness of our results,
we computed the weighted average from the single source fits in each
redshift bin.  The best-fit values resulting from the {\em combined}
fits are always consistent with the weighted means (listed in
Table~\ref{tableZ}) within $1\sigma$, except for the bin at 
$z\sim0.6$, which is lower.  
We also checked that, if the two clusters with
the highest $Z_{Fe}$ (i.e. ZW~0024 and V1416, see Sect.~3.3) are
removed from the fit, the average $Z_{Fe}$ value in the first redshift
bin is only slightly changed (at the level of $\sim3\%$).

\begin{table}
\caption{Average iron abundance in different redshift bins resulting
from combined fit and weighted mean.}
\label{tableZ}
\centering
\begin{tabular}{l l l l}
\hline\hline
$\langle z\rangle\mathrm{^a}$ & $Z/Z_\odot\mathrm{^b}$ & $Z/Z_\odot\mathrm{^c}$ & 
$\Delta Z/Z_\odot\mathrm{^d}$ \\
 & (combined fit) & (weighted mean) & ({\sl rms}) \\
\hline
0.206 [9]     & $0.427^{+0.003}_{-0.011}$ & $0.416 \pm 0.009$  & $0.08$  \\
0.350 [10]    & $0.387_{-0.012}^{+0.013}$ & $0.379 \pm 0.019$  & $0.20$  \\
0.447 [12]    & $0.330^{+0.017}_{-0.012}$ & $0.318 \pm 0.020$  & $0.11$  \\
0.572 [15]    & $0.306^{+0.017}_{-0.033}$ & $0.260 \pm 0.020$  & $0.13$  \\
0.787 [12]    & $0.244\pm0.025$           & $0.251 \pm 0.035$  & $0.15$  \\
1.167 [7]     & $0.265_{-0.04}^{+0.05}$   & $0.28 \pm 0.048$  & $0.15$  \\ 
\hline
\end{tabular}
\begin{list}{}{}
\item[Notes:] $\mathrm{^a}$ average redshift of each bin (the number of clusters 
in each bin is shown in parenthesis);
$\mathrm{^b}$ iron abundance from combined fit with $1\sigma$ errors;
$\mathrm{^c}$ iron abundance from weighted mean; 
$\mathrm{^d}$ {\em rms} dispersion. 
\end{list}
\end{table}

From a visual inspection of Fig.~\ref{Z_vs_z}, we notice a
continuous trend of decreasing iron abundance from $z\simeq0.3$ to
$z\simeq1.2$. While a constant value $\langle Z_{Fe}
\rangle\simeq 0.25 \,Z_\odot$ is a good fit at $z>0.5$, the iron
abundance is significantly higher at $z<0.5$, the redshift range over
which the statistics of our sample increased most with respect to
Paper~I. In addition, we now have a firm measurement of the average
iron abundance at redshift $z\simeq0.8$ and $z\simeq1.2$, reinforcing
the results of Paper~I; in particular, the iron abundance in the most
distant clusters is still consistent with the value
$Z_{Fe}=0.3\,Z_\odot$ within $1\sigma$.

Given the negative correlation between iron abundance and temperature
found in our sample (see Sect.~3.3), we first verified whether the
evolution with redshift is due to the presence of low-temperature
clusters in the low redshift bins. By excluding clusters with
$kT<5$~keV from the plot, the same results are obtained, as expected
from the lack of correlation betwen the average temperature and
redshift found in Sect.~3.2.  

We also note that this trend points towards $\langle Z_{Fe}\rangle
\sim 0.5 \, Z_\odot$ at low-z, which is higher than the often-reported 
value $Z_{Fe} \simeq 0.3\, Z_\odot$.  The reason is that we
are measuring $Z_{Fe}$ in the inner regions of clusters, where it
reaches values significantly higher than $0.3\, Z_\odot$, particularly
in cool-core clusters (see \citealt{vik05}), which constitute about
2/3 of the local population.  In order to show that the average iron
abundance in low-z clusters, when analyzed with our procedure, confirm
the trend seen in the high-z sample, we add a point at $\langle z
\rangle\simeq 0.2$ including 9 clusters, as described in detail in
Sect.~3.5.

\begin{figure}
\centering
\includegraphics[width=8.0 cm, angle=0]{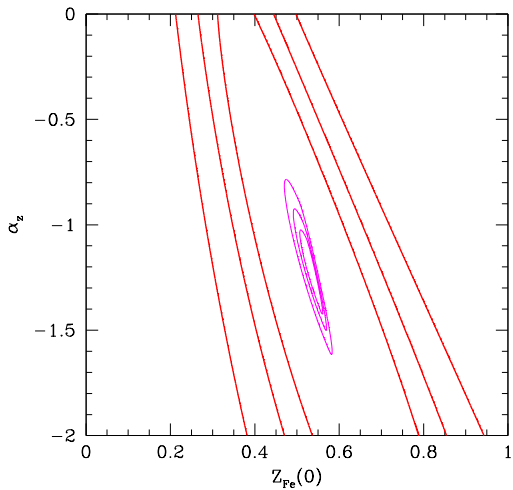}
\caption{Best-fit confidence contours plot for the metallicity as a
function of redshift (including the additional low-z point) modelled
with a power law of the form $Z(z)=Z_{Fe}(0)\,(1+z)^{-\alpha_z}$.
Thick inner contours display the 1, 2, and $3\sigma$ c.l. using the results
of the combined fits, while thin outer contours are obtained using
the {\em rms} dispersion.}
\label{cont_zmet} 
\end{figure}

A fit with a constant iron abundance value over the entire redshift
range is unacceptable.  If we model the evolution (including the
additional low-z point, using the values from the combined fits) with
a simple power law as

\begin{equation}
\langle Z_{Fe}\rangle \simeq Z_{Fe}(0)\,(1+z)^{-\alpha_z} \, ,
\end{equation}
\noindent
the best-fit values obtained are $Z_{Fe}(0)=0.54\pm 0.04\,Z_\odot$ and
$\alpha_z=1.25\pm0.15$ ($\chi_\nu^2=0.9$), implying a decrease by a
factor of $\sim 2$ between $z=0.3$ and $1.2$.  The evolution
($\alpha_z<0$) is significant only at $1\sigma$ level when the {\em
  rms} dispersion is used instead of the errors on the combined fits
(see confidence contours in Fig.~\ref{cont_zmet}). However, the {\em
  rms} dispersion is greatly overestimating the uncertainties on the
average values. Consistent best-fit results are obtained if the
lowest redshift point is excluded from the fit or if weighted mean
values are used.

\subsection{The ``local" iron abundance of the ICM}
 
An important issue to address is how our findings at $z\ga0.3$ compare
with the {\em local} iron abundance, which is traditionally quoted to
be $Z_{Fe}\simeq0.3\,Z_\odot$ \citep{ren97} in the units following Anders \&
Grevesse (1989).  The spatially-resolved spectroscopy of local clusters
obtained with the {\em Chandra} and XMM-{\em Newton} satellites shows
a complex distribution of the metals within the inner regions. The
large differences in iron abundances and gradients from cluster to
cluster lessens the meaning of the adoption of a single canonical value 
for the average iron content of the ICM in the local Universe.
The value $Z_{Fe}\simeq0.3\,Z_\odot$ is typically observed only in non cool-core
clusters or in the outer regions ($>100$~kpc) of cool-core clusters
(see \citealt{tam04} for XMM-{\em Newton} and \citealt{vik05} for {\em
Chandra} data). The central peak of iron abundance reaches values of
$Z_{Fe}\simeq (0.6-0.8)\,Z_\odot$ in the cores of cool-core clusters,
which have a typical size of 100~kpc \citep[see][]{deg04, vik05},
whereas $Z_{Fe}$ decreases to $\simeq 0.3\,Z_\odot$ in the outer
regions. On the other hand, the iron abundance appears to be constant,
$Z_{Fe}\simeq 0.2-0.3\,Z_\odot$, in non cool-core clusters.  As a
result, particular care should be used when comparing our measurements
with the local values of $Z_{Fe}$ from the literature.

Since the extrapolation of the average $Z_{Fe}$ at low-z points
towards $Z_{Fe}(0) \simeq 0.5\, Z_\odot$, we need to explain the
apparent discrepancy with the oft-quoted canonical value $\langle
Z_{Fe}\rangle \simeq 0.3\, Z_\odot$.  As mentioned in Sect.~3.4, the
discrepancy is due to the fact that our average values are computed
within $r\simeq 0.15\, R_{vir}$, where the iron abundance is boosted
by the presence of metallicity peaks often associated to cool cores.
The regions chosen for our spectral analysis, are larger than the
typical size of the cool cores, but smaller than the typical regions
adopted in studies of local samples.

This can be proved by analyzing the inner regions ($r<0.15\, R_{vir}$)
of a sample of clusters at $z<0.3$. We selected a small subsample of 9
clusters at redshift $0.1<z<0.3$, including 7 cool-core and 2 non
cool-core clusters, a mix that is representative of the low-z
population.  These clusters, listed in Table~\ref{lowz}, are presently
being analyzed for a separate project aimed at obtaining spatially-resolved 
spectroscopy (Baldi et al., in preparation). Here we analyze
a region within $r=0.15\, R_{vir}$ in order to probe the same regions
probed at high redshift.  We used this small control sample to add a
low-redshift point in our Fig.~\ref{Z_vs_z}, which extends the
$Z_{Fe}$ evolutionary trend.

\begin{table*}
\caption{Spectral fit results for the low-z sample with the {\tt
tbabs(mekal)} model.}
\label{lowz}
\centering
\begin{tabular}{l l l l l l l}
\hline\hline
Cluster & z & kT [keV]$\mathrm{^a}$ & $Z/Z_\odot\mathrm{^b}$ & 
$N_H$ [cm$^{-2}$]$\mathrm{^c}$ & $\chi^2_r$ [d.o.f.]$\mathrm{^d}$ & Null Hyp.
Prob.$\mathrm{^e}$ \\
\hline
\object{Abell~1413}         &   0.143 &  $7.1 \pm 0.1$      &  $0.39\pm  0.02$            & $2.18\times10^{20}$ & 1.55 [477]   &   $10^{-13}$ \\
\object{Abell~907}          &   0.153 &  $5.2 \pm 0.1$      &  $0.53  \pm 0.03$         & $5.36\times10^{20}$ & 1.37 [431]   &   $10^{-7}$ \\
\object{Abell~2104}         &   0.155 &  $13.9\pm 0.5 $      &  $0.53_{-0.07}^{+0.04}$  & $8.69\times10^{20}$ & 1.35 [423]   &   $10^{-6}$ \\
\object{Abell~2218}         &   0.176 &  $7.9\pm 0.3 $      &  $0.26 \pm 0.03 $         & $3.26\times10^{20}$ & 1.09 [347]   &   0.111   \\
\object{Abell~963}          &   0.206 &  $7.0\pm 0.2$      &  $0.43 \pm  0.03$         & $1.40\times10^{20}$ & 1.09 [341]   &   0.113   \\
\object{Abell~2261}         &   0.224 &  $7.5_{-0.2}^{+0.4}$ &  $0.51_{-0.05}^{+0.03}$  & $3.28\times10^{20}$ & 1.07 [329]   &   0.169   \\
\object{Abell~2390}         &   0.228 &  $9.1 \pm  0.1$     &  $0.40_{-0.02}^{+0.03}$  & $6.81\times10^{20}$ & 1.50 [479]   &   $10^{-12}$ \\
\object{Abell~1835}         &   0.253 &  $7.2 \pm 0.2  $     &  $0.41_{-0.04}^{+0.03}$  & $2.32\times10^{20}$ & 0.99 [291]   &   0.527   \\
\object{ZwCl~$1021.0+0426$} &   0.291 &  $6.2\pm 0.1 $     &  $0.39_{-0.03}^{+0.03}$  & $3.02\times10^{20}$ & 1.51 [396]   &   $10^{-10}$ \\
\hline
\end{tabular}
\begin{list}{}{}
\item[Notes:] $\mathrm{^a}$ temperature; $\mathrm{^b}$ iron abundance in solar units 
by \citet{and89}; $\mathrm{^c}$ local column 
density, always fixed to the Galactic value by Dickey \& Lockman (1990); 
$\mathrm{^d}$ reduced chi-square and degrees of freedom obtained
after binning the spectra to 20 counts per bin; $\mathrm{^e}$ null-hypothesis 
probability. Errors refer to the $1\sigma$ confidence level. 
\end{list}
\end{table*}

\section{Discussion}

The main result of this work is that the cosmic average of $Z_{Fe}$ in
the central regions ($R< 0.3\,R_{vir}$) of clusters significantly
decreases with redshift out to $z \simeq 0.5$, remaining constant out
to $z \simeq 1.3$ at the level of $Z_{Fe} \simeq 0.25$.  Given the
complex thermal and chemical structures observed in bright local
clusters, a main concern is whether our analysis might be affected by
evolution in the occurrence of temperature/metallicity gradients in
the cluster population. The assumption of a single-temperature 
{\tt mekal} model for the inner $0.3\,R_{vir}$ may well be too 
simplistic, and it may introduce systematic biases in the
recovered $Z_{Fe}$ values. In order to clarify this issue, we
investigate possible biases in our fitting procedure in the Appendix
using a large set of simulated spectra in the typical S/N regime of
our high-z clusters.  We find that different S/N values do not
introduce any significant bias. In particular, spectra with lower S/N
(occurring mostly at high-redshift) tend to give slightly higher
$Z_{Fe}$ compared to the input values, therefore opposite
to the observed trend (see Fig.~A.1).

In the Appendix we also investigate the cases of a two-temperature
ICM with a single $Z_{Fe}$ and of a two-temperature ICM with higher
$Z_{Fe}$ associated with the colder component, analyzed with a
single-temperature {\tt mekal} model.  Here the key parameter is the
ratio of the emission measure of the two components.  This quantity is
difficult to model, so that here we assume a few representative
values ranging from 0.3 to 0.75.

In both cases we measured slightly higher temperatures at higher
redshifts, due to the fact that for high-redshift clusters, the
signature of the cold component is partially redshifted below the
adopted energy range ($E>0.6$~keV).  We also find a mild trend toward
lower $Z_{Fe}$ at $z\sim 1$ compared to $z\sim 0.6$.  This effect
is limited to be $\leq 30$\%, and therefore it cannot fully explain
the observed decrease even under the extreme assumption that {\em all}
the clusters in the sample had steep temperature gradients with a
central abundance peak.  To sum up, these simulations show that the
presence of temperature and abundance gradients, {\em if their
occurrence is constant through the population of clusters at different
redshifts}, does not introduce significant bias into our measure of the
evolution of $Z_{Fe}$.

Furthermore, we investigate whether the evolution of $Z_{Fe}$ could be
due to an evolving fraction of clusters with cool cores, which are
known to be associated with iron-rich cores \citep[see][]{deg04} and
which amount to more than 2/3 of the local clusters \citep[see][]{bau05}.
For example, the evolution of the mass in iron in the central peak,
which is about $20-30$\% of the total, may be associated with the star
formation product of the central galaxy alone \citep[see][]{deg04}.

\begin{figure}
 \centering
 \includegraphics[width=7.5 cm, angle=0]{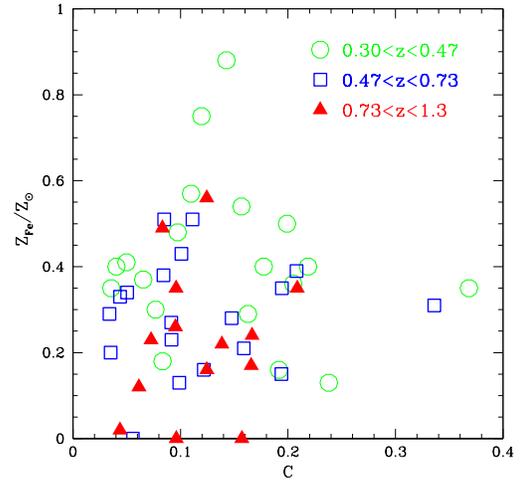}
 \caption{Iron abundance plotted versus $C = f(r<50\,kpc)/f(r<500\,kpc)$.
 Clusters within different redshift bins are coded with different symbols.}
 \label{sbr_vs_z}
 \end{figure}

In order to use a simple characterization of cool-core clusters
in our high-z sample, we computed the ratio of the fluxes emitted
within 50 and 500~kpc ($C = f(r<50\,kpc)/f(r<500\,kpc$) computed as
the integral of the surface brightness in the $0.5-5$~keV band
(observer frame).  This quantity ranges between 0 and 1 and it
represents the relative weight of the central surface brightness.
Higher values of $C$ are expected if a cool core is present.  If the
decrease in $Z_{Fe}$ with redshift is associated to a decrease in the
number of cool-core clusters for higher $z$, we would expect to
observe a positive correlation between $Z_{Fe}$ and $C$ and a
negative correlation between $C$ and redshift.  In
Fig.~\ref{sbr_vs_z} we plot $Z_{Fe}$ as a function of $C$ for our
sample.  We find that in our sample there is no correlation between
metallicity and $C$ with a Spearman's coefficient of $r_s =0.02$
(significance of $\sim 0.2 \sigma$) nor one between $C$ and redshift
($r_s=-0.11$, a level of confidence of $0.8 \sigma$).

The absence of strong correlations between $C$ and iron abundance or
between $C$ and redshift suggests that the mix of cool cores and
non cool cores over the redshift range studied in the present work
cannot justify the observed evolution in the iron abundance.
We caution, however, that a possible evolution of the occurrence of
cool-core clusters at high redshift may still partially contribute to
the observed evolution of $Z_{Fe}$. In other words, whether the
observed evolution of $Z_{Fe}$ is contributed entirely by the
evolution of the mass of iron or is partially due to a
redistribution of iron in the central regions of clusters is an open
issue to be addressed with a proper and careful investigation
of the surface brightness of the high-z sample.

\begin{figure}
\centering
\includegraphics[width=7.5 cm, angle=0]{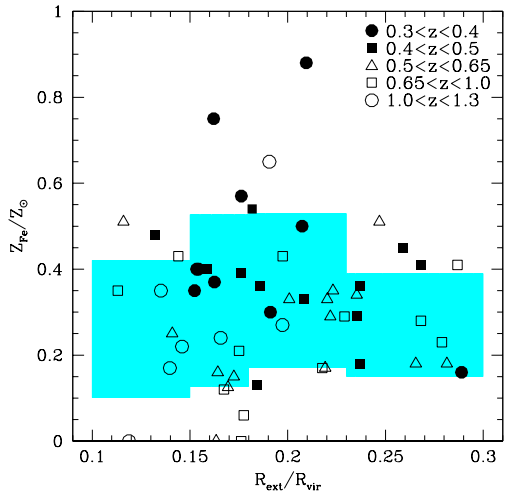}
\caption{Iron abundance plotted versus the ratio $R_{ext}/R_{vir}$.
Shaded areas show the {\em rms} dispersion around the average iron
abundance in four bins.}
\label{zfe_vs_rext}
\end{figure}

A final check is provided by the scatter plot of $Z_{Fe}$ versus
$R_{ext}/R_{vir}$, shown in Fig.~\ref{zfe_vs_rext}.  We do not
detect any dependence of $Z_{Fe}$ on the extraction radius adopted for
the spectral analysis.  In particular, we find that clusters with
smaller extraction radii do not show higher $Z_{Fe}$ values.

Together with the previous discussion of possible selection biases of
our sample, all these tests concur to indicate that the observed
evolution of the iron abundance is a genuine signature of some
physical processes associated with the production and release of iron
into the ICM.

This finding may be directly interpreted in terms of the cosmic star
formation history \citep[see][]{ett05} for high assumed values of the
time delay of SNe~Ia \citep[see][]{dah04}, which are expected to be
the main contributors of iron. Other works \citep[e.g.][]{man06} argue
that the data on (i) the evolution of the SN~Ia rate with redshift, (ii)
the enhancement of the SN~Ia rate in radio-loud early type galaxies,
and (iii) the dependence of the SN~Ia rate on the colours of the
parent galaxies suggest the existence of two populations of
progenitors for SN~Ia. One population is expected to explode soon
after the stellar birth on a time scale of $10^8$ years and can
significantly pollute the ICM with iron at high redshift. The second
population contributes to late enrichment with an exponential decay
time of $\sim3$~Gyrs. Following the argument described in
\citet{ett05}, we note that the rates adopted by \citet{man05} predict
a flatter distribution of the iron abundance as function of redshift
than the rates tabulated in \citet{dah04}, with a milder negative
evolution still partially consistent with our measurements.  Using
detailed chemical evolution models, \citet{low06} has recently
interpreted the significant enrichment of the ICM at $z\simeq1$ as
direct evidence of prompt star formation in spheroids with a top-heavy
IMF.  Moreover, these models generally predict a significant increase
in the iron abundance between $z=0$ and $z=1$, in qualitative agreement
with our results.

Alternatively, the evolution in $Z_{Fe}$ occurring in the 5~Gyr
spanned by our sample may be ascribed to some dynamical processes
that transfer the metal-enriched gas from the intergalactic medium of
the cluster galaxies to the hot phase of the ICM.  Mechanisms such as
ram pressure \citep[][]{dom04,cor06} or tidal stripping
\citep[e.g.][]{mur04} are currently being investigated with numerical
simulations \citep[see also][]{val03,tor04}. The same mechanisms are
often invoked to explain the morphological evolution of the cluster
galaxies' population.

\section{Conclusions}

We have presented the spectral analysis of 56 clusters of galaxies at
intermediate-to-high redshifts observed by {\em Chandra} and XMM-{\em
Newton}. This work improves our first analysis aimed at tracing the
evolution of the iron content of the ICM out to $z\ga1$ (Paper~I), by
substantially extending the sample. The main results of our work can
be summarized as follows:

\begin{itemize}

\item We determine the average ICM iron abundance with a $\sim20$\%
uncertainty at $z>1$ ($Z_{Fe}=0.27\pm0.05\,Z_\odot$), thus confirming
the presence of a significant amount of iron in high-z
clusters. $Z_{Fe}$ is constant above $z\simeq0.5$, the largest
variations being measured at lower redshifts.

\item We find a significantly higher average iron abundance in
clusters with $kT<5$~keV, in agreement with trends measured in local
samples. For $kT>3$~keV, $Z_{Fe}$ scales with temperature as
$Z_{Fe}(T)\simeq0.88\,T^{-0.47}$.

\item We find significant evidence of a decrease in $Z_{Fe}$ as a
function of redshift, which can be parametrized by a power law $\langle
Z_{Fe}\rangle \simeq Z_{Fe}(0)\,(1+z)^{-\alpha_z}$, with
$Z_{Fe}(0)\simeq0.54 \pm 0.04$ and $\alpha_z\simeq1.25 \pm 0.15$.
This implies an evolution of more than a factor of 2 from $z=0.4$ to
$z=1.3$.

\end{itemize}

We carefully checked that the extrapolation towards $z \simeq 0.2$ of
the measured trend, pointing to $Z_{Fe} \simeq 0.5\, Z_\odot$, is
consistent with the values measured within a radius $r= 0.15\,
R_{vir}$ in local samples including a mix of cool-core and non
cool-core clusters. We also investigated whether the observed
evolution is driven by a negative evolution in the occurrence of
cool-core clusters with strong metallicity gradients towards the
center, but we do not find any clear evidence of this effect. We
note, however, that a proper investigation of the thermal and
chemical properties of the central regions of high-z clusters is
necessary to confirm whether the observed evolution by a factor of
$\sim2$ between $z=0.4$ and $z=1.3$ is due entirely to physical
processes associated with the production and release of iron into the
ICM, or partially associated with a redistribution of metals connected
to the evolution of cool cores.

Precise measurements of the metal content of clusters over large
look-back times provide a useful fossil record for the past star
formation history of cluster baryons.  A significant iron abundance in
the ICM up to $z\simeq 1.2$ is consistent with a peak in star
formation for proto-cluster regions occurring at redshift
$z\simeq4-5$. On the other hand, a positive evolution of $Z_{Fe}$ with
cosmic time in the last 5~Gyrs is expected on the basis of the
observed cosmic star formation rate for a set of chemical enrichment
models.  Our data provide further constraints on the chemical
evolution of cosmic baryons in the hot diffuse and cold phases.

\begin{acknowledgements}
P. Tozzi acknowledges support under the ESO visitor program in
Garching during the completion of this work (April--May 2004;
May--June 2006).  The authors are deeply indebted to A. Baldi for
providing the data of the low-z sample before publication.  We thank
A. Vikhlinin and N. Cappelluti for helpful discussion of the
reduction and spectral analysis of {\em Chandra} data. We thank
S. De~Grandi and S. Molendi for helpful discussions. We
acknowledge the financial contribution from contract ASI--INAF I/023/05/0
and from the PD51 INFN grant.
We are grateful to the anonymous referee for the valuable comments 
and suggestions.
\end{acknowledgements}

\bibliographystyle{aa}
\bibliography{sbs}

\appendix

\section{Analysis of simulated X-ray spectra}

In this section we investigate the presence of a possible bias in the
measure of $Z_{Fe}$ in our analysis procedure.  In particular, we
check whether unresolved gradients in the temperature or in the iron
abundance distribution can affect the observed trends, paying
particular attention to the low S/N regime of our spectra.  We perform
several simulations of spectra with different assumptions, as
described in detail in the following subsections, and explore the
possible conditions that can potentially affect the distribution of
best-fit values of $kT$ and, most important, of $Z_{Fe}$.  The median
of the distribution of best-fit values and the 16\% and 84\%
percentiles (corresponding to the $1\sigma$ confidence level) will
finally be compared to the input values of temperature and metallicity.

\subsection{Fitting bias in isothermal, constant metallicity, 
low S/N spectra}

A first simple test is to check the accuracy that we can achieve in
recovering the input parameters of temperature and metallicity from
spectra simulated with the typical S/N, temperatures and redshifts of
clusters in our sample, under the assumption of a single-temperature
{\tt mekal} model.  This may seem a redundant exercise.  However a
potential problem rises from the fact that upper limits on
temperature are typically less constrained than lower limits.  This
effect increases at high temperatures, when the exponential cut off of
the thermal spectrum shifts to energies for which the effective area
of the detectors is low.  As a consequence, if temperature best-fit
values tend to be scattered upwards for low S/N, the estimated
continuum may be higher than the actual one, and therefore the
measured equivalent width of the iron lines may be underestimated. On
the other hand, variations in the best-fit temperatures affect the ion
abundances too.  Spectral simulations can be used to investigate how
these aspects affect the measure of $Z_{Fe}$ at low S/N as a function
of redshift.  

In principle, the parameter space to explore is fairly wide: input
metallicity, input temperature, redshifts, and S/N (measured as $S/N
\equiv C_s /\sqrt{C_{tot}+C_{bck}}$, where $C_s$ is the number of net
counts from the source, $C_{tot}$ the number of source plus
background counts, and $C_{bck}$ the number of estimated background
counts in the source area). For simplicity, we have decided to restrict the
simulations to a few relevant cases. We chose an input metallicity of
$Z_{Fe}=0.3\,Z_\odot$, redshift $z=0.4$, and temperatures of $kT=3.5$
and 7~keV. We took the observation of V~1416 as a template:
the exposure time is 30~ks and the extraction radius $74\arcsec$. The
simulations were performed for four different values of S/N (corresponding
to 1850, 1000, 500, and 200 net counts). We also run two simulations
for $z=1$. Each combination of parameters was simulated 1000
times. The simulations were performed with XSPEC using a {\tt mekal}
model. We analyzed each simulated spectrum by adopting the same input
model. In Table~\ref{simtab1} we list the median of the distributions
of the best-fit values for temperature and $Z_{Fe}$. Errors on the
median are the 16\% and 84\% percentiles.

\begin{table*}
\caption{Results from the spectral simulations described in Appendix A.1.}
\label{simtab1}
\centering
\begin{tabular}{l l l l l l l l }
\hline\hline
Sim$\mathrm{^a}$ & net cts$\mathrm{^b}$ & S/N$\mathrm{^c}$ & $kT_{inp}\mathrm{^d}$ & 
$Z_{inp}\mathrm{^e}$ & $z\mathrm{^f}$ & $kT_{fit}\mathrm{^g}$ & $Z_{fit}\mathrm{^h}$ \\
\hline
s01 & 1855 & 34.4 & 3.5 & 0.3 & 0.4 & $3.66^{+0.27}_{-0.28}$ & $0.33^{+0.15}_{-0.15}$ \\ 
s02 & 1000 & 22   & 3.5 & 0.3 & 0.4 & $3.82^{+0.48}_{-0.41}$ & $0.36^{+0.31}_{-0.22}$ \\ 
s03 & 500  & 12   & 3.5 & 0.3 & 0.4 & $4.15^{+1.01}_{-0.63}$ & $0.44^{+0.46}_{-0.42}$ \\ 
s04 & 200  & 5.6  & 3.5 & 0.3 & 0.4 & $5.5^{+6.5}_{-1.9}$    & $0.44^{+1.52}_{-0.44}$ \\ 
s05 & 1920 & 35.2 & 7   & 0.3 & 0.4 & $7.5^{+1.0}_{-0.9}$    & $0.32^{+0.18}_{-0.19}$ \\ 
s06 & 990  & 22   & 7   & 0.3 & 0.4 & $7.95^{+2.17}_{-2.49}$ & $0.32^{+0.31}_{-0.28}$ \\ 
s07 & 520  & 13   & 7   & 0.3 & 0.4 & $9.45^{+5.75}_{-2.65}$ & $0.40^{+0.61}_{-0.40}$ \\ 
s08 & 330  & 8.9  & 7   & 0.3 & 0.4 & $11.8_{-5.3}^{+18.7}$  & $0.55^{+1.23}_{-0.55}$ \\ 
s09 & 1000 & 22   & 7   & 0.3 & 1.0 & $8.07_{-1.39}^{+1.78}$ & $0.29^{+0.27}_{-0.28}$ \\ 
s10 & 500  & 12.7 & 7   & 0.3 & 1.0 & $9.0_{-2.1}^{+4.0}$    & $0.28^{+0.62}_{-0.28}$ \\ 
\hline
\end{tabular}
\begin{list}{}{}
\item[Notes:] $\mathrm{^a}$ simulations identification number; 
$\mathrm{^b}$ net number of counts; 
$\mathrm{^c}$ signal-to-noise ratio;
$\mathrm{^d}$ input temperature;
$\mathrm{^e}$ input iron abundance;
$\mathrm{^f}$ redshift;
$\mathrm{^g}$ median values of the distribution of best-fit temperatures;
$\mathrm{^h}$ median values of the distribution of best-fit iron abundances. Lower and
upper errors correspond to the 16\% and 84\% percentiles, respectively. 
\end{list}
\end{table*}
 
\begin{figure}
\centering
\includegraphics[width=7.0 cm, angle=0]{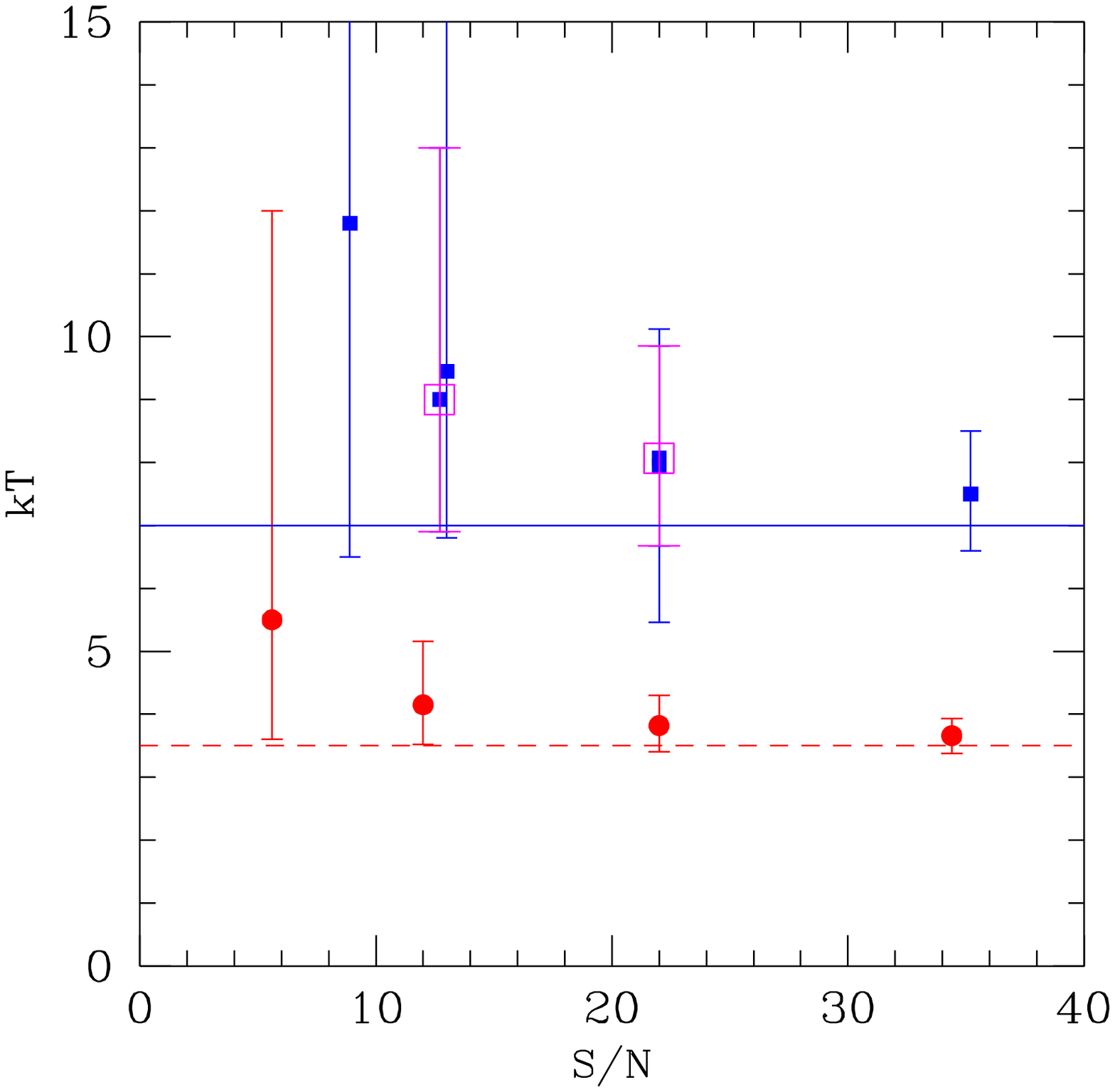}
\includegraphics[width=7.0 cm, angle=0]{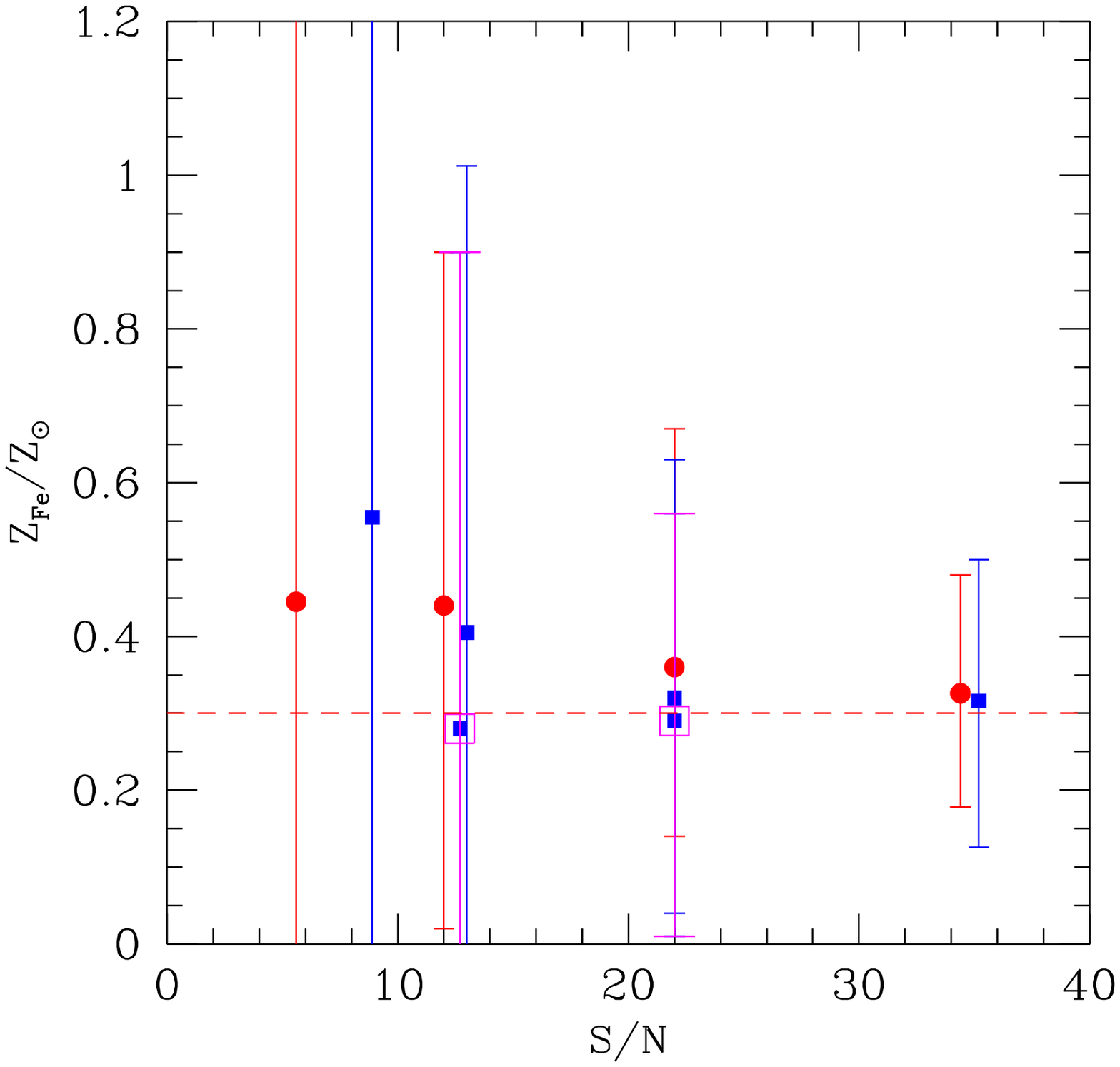}
\caption{{\em Upper Panel:} Median of the best-fit temperature 
distribution as a function of S/N. Red solid circles refer to an input
temperature value of 3.5~keV (dashed horizontal line), while blue
solid squares to 7~keV (solid horizontal line). Empty squares are for
$z=1$. Lower and upper error bars correspond to the 16\% and the 84\%
percentiles, respectively. {\em Lower Panel:} Median of the best-fit 
$Z_{Fe}$ distribution as a function of S/N. The input value is
always $Z_{Fe}=0.3\,Z_\odot$ (horizontal dashed line). Symbols and
error bars are as in the upper panel; magenta empty squares are for $z=1$.}
\label{simul1}
\end{figure}

The results are summarized in Fig.~\ref{simul1}. We notice that, as
expected, the low S/N spectra tend to have higher median best-fit
temperatures compared to the input values. This translates into a
slightly higher median of the best-fit $Z_{Fe}$ than the
input value, which is always $Z_{Fe}=0.3\,Z_\odot$ for this set of
simulations. However, for $S/N < 20$, the distribution of best-fit
values is largely scattered around the input values. At higher
redshifts, the situation for a given S/N improves slightly, since the
exponential cutoff moves towards the most sensitive energy range of
{\em Chandra}, and both the temperature and the abundance estimates
are closer to the input values. We conclude that, under the assumption
of a single temperature, single metallicity thermal plasma, the
best-fit values of $Z_{Fe}$ are not significantly biased, within the 
typical S/N and redshift range of our sample.

\subsection{Fitting bias in two-temperature, constant-metallicity 
spectra}

Here we intend to investigate whether the presence of substantial temperature
structure can affect the measure of the iron abundance when the
spectra are analyzed with a single-temperature {\tt mekal} model, as adopted
in this work. In some cases in our spectra, we are able to detect the
presence of a temperature decrease towards the center; however, we are
not able to perform a spatially-resolved spectral analysis for the
large majority of clusters in our sample.

\begin{figure}
\centering
\includegraphics[width=7.0 cm, angle=0]{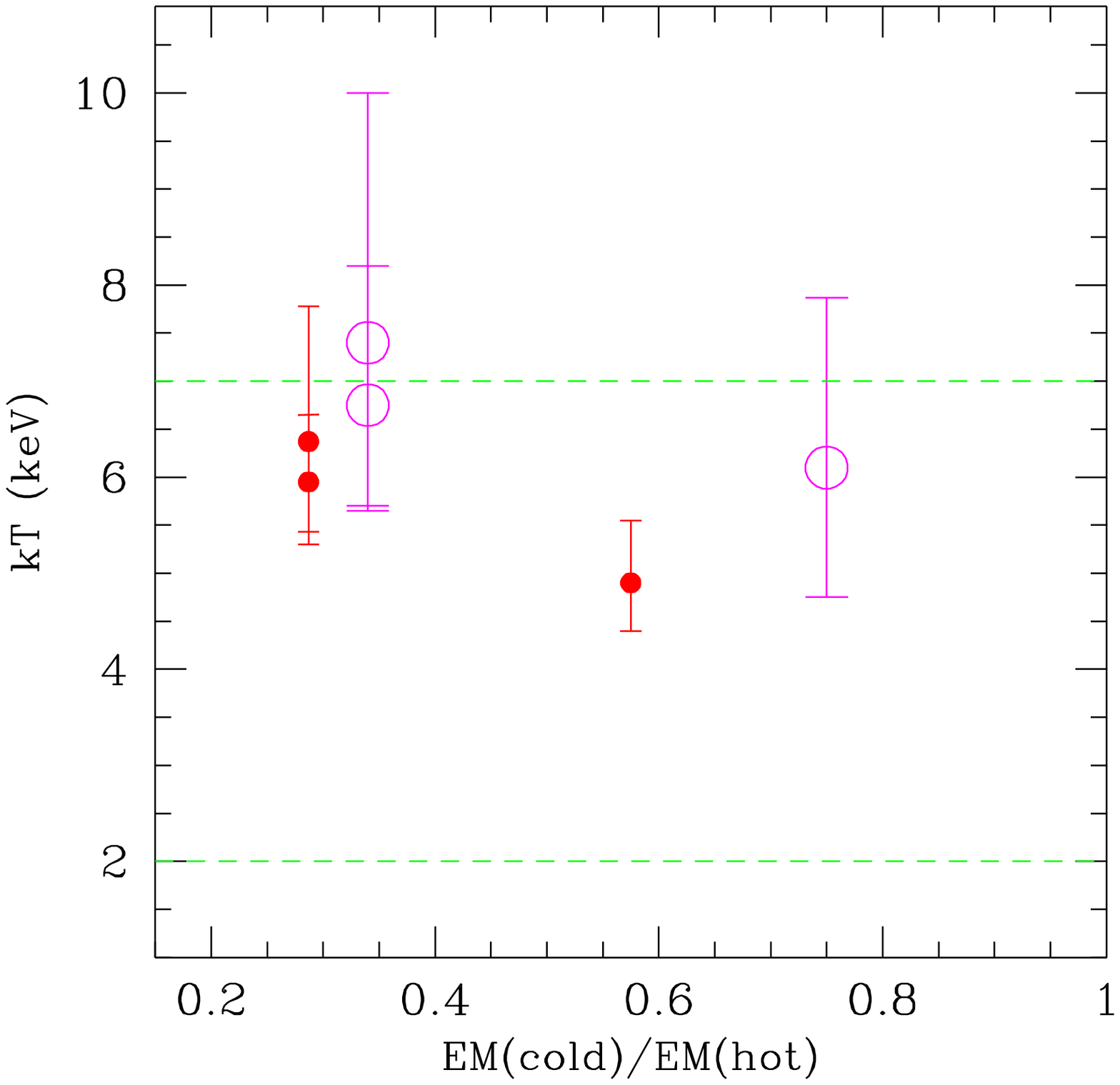}
\includegraphics[width=7.0 cm, angle=0]{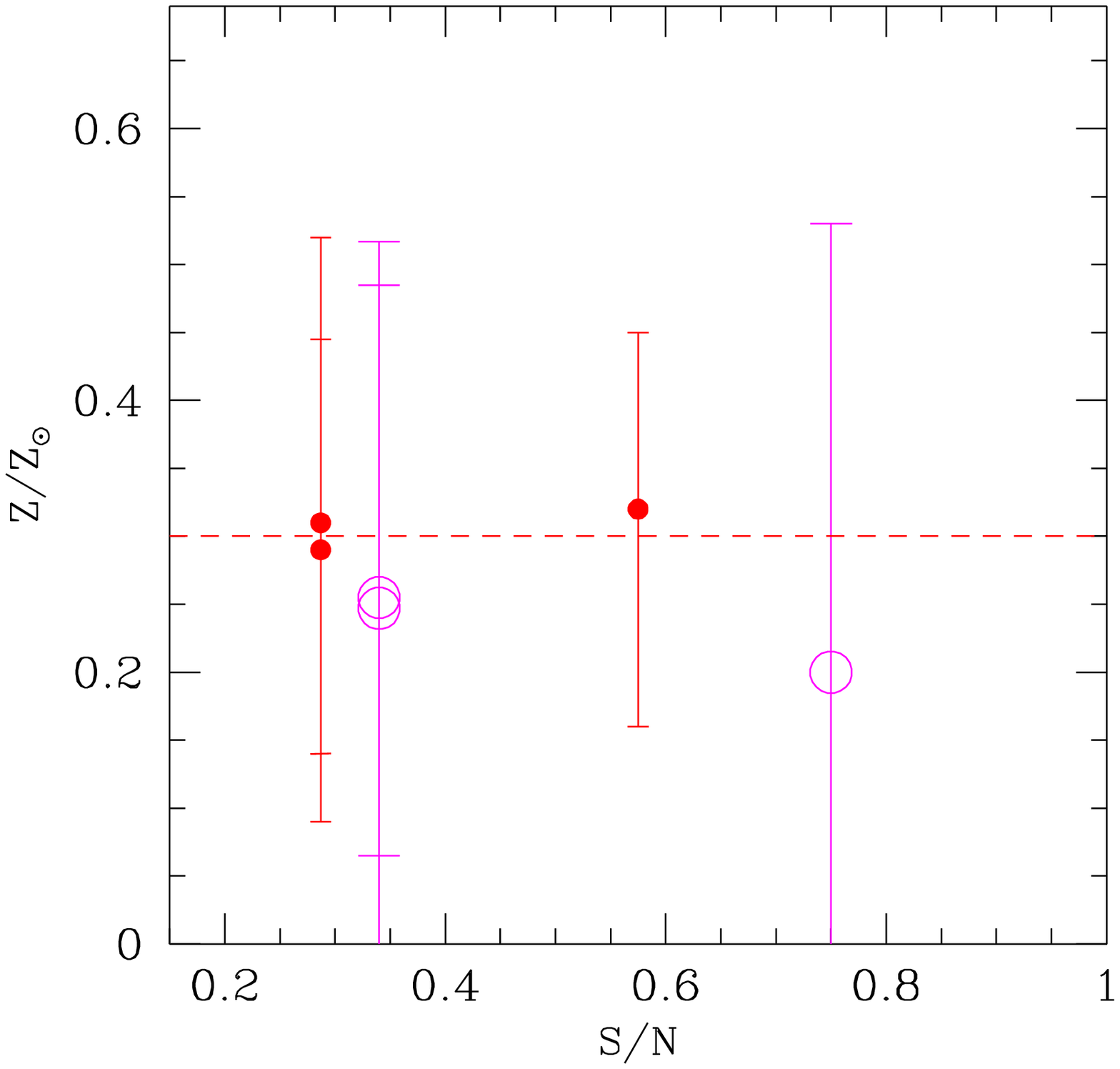}
\caption{{\em Upper Panel:} median values of the best-fit temperature 
distribution as a function of the ratio of the cold and hot emission measure.
Input temperatures are 2 and 7~keV (horizontal dashed lines). Filled
circles are for $z=0.6$, while empty circles are for $z=1$. Lower and
upper error bars correspond to the 16\% and 84\% percentiles,
respectively. {\em Lower Panel:} median values of the best-fit $Z_{Fe}$
distribution as a function of the ratio of the cold and hot emission
measure. Symbols and errors as in the upper panel.}
\label{simul2}
\end{figure}

Since there are no canonical, physically-motivated, multiphase models
of the ICM, we simply assume a double {\tt mekal} model. Again, the
free parameter space is wide, therefore here we only explore a few
cases. In particular, we performed a set of simulations of a cluster
with two temperature components (2 and 7~keV), with an emission
measure ratio (EM, the normalization of the {\tt mekal} model) of the
cold to the hot component ranging from 0.3 to 0.75. The choice of the
temperature range here agrees with the observational
evidence that the minimum temperature in cool cores is always equal to
or higher than one third of the maximum value of the hot component
\citep[see][]{pet03}. The iron abundance is set to $0.3\,Z_\odot$ in
both components. We simulated 1000 spectra for each case at redshift
$z=0.6$ and $z=1$.  We took the observation of RX~J0542 as a template:
the exposure time is 50~ks and the extraction radius is $79\arcsec$.
The results are listed in Table~\ref{simtab2}, and shown in 
Fig.~\ref{simul2}. Obviously, for a given redshift, the best-fit
temperature, which is always intermediate between the two input
values, moves towards the low value for increasing values of the cold
to hot EM ratio. But, for a given EM value, the median of the best-fit
temperature moves towards higher values for higher redshifts, since
the cold components is redshifted out of the adopted energy range
($0.6-8$~keV). We note that, despite a two-temperature structure
being fitted with a single-temperature, the fits are acceptable in the
selected cases, due to the low number of total net counts (below
2000), representing the typical condition under which the presence of
two temperatures cannot be established from the spectral analysis.

As for $Z_{Fe}$, the median of the distribution of the best-fit values
is slightly lower at $z=1$ than at $z=0.6$. The effect is a decrement
of less than 30\% up to $z=1$.  Given the large dispersion of the
best-fit values, this effect is probably not playing a dominant role in
our observed trend.  However, a proper investigation of the effect
of a multi-temperature ICM must rely on a physical modelization of
the ICM multiphase structure, which is presently missing.

\begin{table*}
\caption{Results from the spectral simulations of two temperature 
components, analyzed with a single-temperature {\tt mekal} model, as
described in Appendix A.2.}
\label{simtab2}
\centering
\begin{tabular}{l l l l l l l l l }
\hline\hline
Sim$\mathrm{^a}$ & net counts (H+C)$\mathrm{^b}$ & EM ratio$\mathrm{^c}$ & 
S/N$\mathrm{^d}$ & $kT_{inp}\mathrm{^e}$ & $Z_{inp}\mathrm{^f}$ & $z\mathrm{^g}$ & 
$kT_{fit}\mathrm{^h}$ & $Z_{fit}\mathrm{^i}$ \\
\hline
s11 & 1560+390 & 0.575 & 30.3 & 7+2 & 0.3 & 0.6 & $4.9_{-0.5}^{+0.65}$  & $0.32^{+0.13}_{-0.16}$ \\ 
s12 & 1410+540 & 0.287 & 30.3 & 7+2 & 0.3 & 0.6 & $5.95_{-0.65}^{+0.7}$ & $0.29^{+0.16}_{-0.15}$ \\ 
s13 & 918+175  & 0.287 & 19.1 & 7+2 & 0.3 & 0.6 & $6.4_{-0.9}^{+1.4}$   & $0.31^{+0.21}_{-0.22}$ \\ 
s14 & 870+170  & 0.34  & 19.8 & 7+2 & 0.3 & 1.0 & $6.75_{-1.1}^{+1.45}$ & $0.26^{+0.23}_{-0.19}$ \\ 
s15 & 550+105  & 0.34  & 12.3 & 7+2 & 0.3 & 1.0 & $7.4_{-1.7}^{+2.6}$   & $0.25^{+0.27}_{-0.25}$ \\ 
s16 & 580+200  & 0.75  & 12.7 & 7+2 & 0.3 & 1.0 & $6.1_{-1.4}^{+1.8}$   & $0.20^{+0.33}_{-0.20}$ \\ 
\hline
\end{tabular}
\begin{list}{}{}
\item[Notes:] $\mathrm{^a}$ simulations identification number; 
$\mathrm{^b}$ net number of counts of the hot and cold component; 
$\mathrm{^c}$ emission measure ratio;
$\mathrm{^d}$ signal-to-noise ratio;
$\mathrm{^e}$ input temperature of the hot and cold component;
$\mathrm{^f}$ input iron abundance;
$\mathrm{^g}$ redshift;
$\mathrm{^h}$ median values of the distribution of best-fit temperatures;
$\mathrm{^i}$ median values of the distribution of best-fit iron abundances. Lower and
upper errors correspond to the 16\% and 84\% percentiles, respectively. 
\end{list}
\end{table*}

\subsection{Fitting bias in low S/N spectra with temperature and 
metallicity gradients}

We investigate the distribution of best-fit values for $Z_{Fe}$ in the
case of an ICM having significant temperature and metallicity
structures at the same time. In particular, we investigate the most
common case of higher $Z_{Fe}$ associated with lower temperature
components, as found in the cool cores. We repeated the simulations
performed in Appendix A.2, by assigning an iron abundance of
$0.6\,Z_\odot$ to the cold component. The results are listed in
Table~\ref{simtab3}, and shown in Fig.~\ref{simul3}.  The best-fit
temperature has the same behavior as described in Appendix A.2.  For
$Z_{Fe}$ as well the situation is analogous to the previous case, with
the median values slightly biased towards higher values.  From a
visual inspection of Figs.~\ref{simul2} and \ref{simul3}, we
notice that the average best-fit values are lower for $z=1$ compared
with $z=0.6$, by an amount on the order of 30\%.  On the
basis of this result, we might expect that the presence of cool,
metal-rich cores, could mimick the observed negative evolution from
$Z_{Fe}\simeq 0.4\,Z_\odot$ at $z\sim 0.3$ to $Z_{Fe}\simeq
0.2\,Z_\odot$ at $z\sim 1.3$ without an effective decrease in the
amount of iron in the central regions.  However, this effect would
explain only part of the observed trend even in the extreme case in
which all the clusters in the sample had strong temperature and
metallicity gradients.  Therefore, we can conclude that the observed
evolution of $Z_{Fe}$ cannot be ascribed entirely to K-correction
effects.  We also notice that the limited effect of a central cold and
metal-rich component in high-z clusters, also implies that the
$Z_{Fe}$-temperature correlation cannot be simply explained by the
occurrence of cool cores with temperature below 2 keV in clusters
with virial temperatures $kT \leq 5 $~keV.

\begin{table*}
\caption{Results from the spectral simulations of two components with
different temperatures and iron abundances, analyzed with a
single-temperature {\tt mekal} model, as described in Appendix A.3.}
\label{simtab3}
\centering
\begin{tabular}{l l l l l l l l l }
\hline\hline
Sim$\mathrm{^a}$ & net counts (H+C)$\mathrm{^b}$ & EM ratio$\mathrm{^c}$ & 
S/N$\mathrm{^d}$ & $kT_{inp}\mathrm{^e}$ & $Z_{inp}\mathrm{^f}$ & $z\mathrm{^g}$ & 
$kT_{fit}\mathrm{^h}$ & $Z_{fit}\mathrm{^i}$ \\
\hline
s17 & 1410+650 & 0.575 & 31.6 & 7+2 & 0.3+0.6 & 0.6 & $4.41_{-0.37}^{+0.46}$ & $0.41^{+0.15}_{-0.14}$ \\ 
s18 & 1640+370 & 0.287 & 31   & 7+2 & 0.3+0.6 & 0.6 & $5.48_{-0.60}^{+0.67}$ & $0.35^{+0.15}_{-0.14}$ \\ 
s19 & 920+210  & 0.287 & 19.5 & 7+2 & 0.3+0.6 & 0.6 & $6.0_{-0.1}^{+1.2}$    & $0.33^{+0.25}_{-0.21}$ \\ 
s20 & 870+200  & 0.34  & 18.8 & 7+2 & 0.3+0.6 & 1.0 & $6.4_{-0.9}^{+1.3}$    & $0.25^{+0.22}_{-0.20}$ \\ 
s21 & 550+125  & 0.34  & 12.6 & 7+2 & 0.3+0.6 & 1.0 & $7.25_{-1.7}^{+2.55}$  & $0.26^{+0.35}_{-0.26}$ \\ 
s22 & 480+240  & 0.75  & 13.3 & 7+2 & 0.3+0.6 & 1.0 & $5.51_{-1.1}^{+2.5}$   & $0.29^{+0.28}_{-0.28}$ \\ 
\hline
\end{tabular}
\begin{list}{}{}
\item[Notes:] $\mathrm{^a}$ simulations identification number; 
$\mathrm{^b}$ net number of counts of the hot and cold component; 
$\mathrm{^c}$ emission measure ratio;
$\mathrm{^d}$ signal-to-noise ratio;
$\mathrm{^e}$ input temperature of the hot and cold component;
$\mathrm{^f}$ input iron abundance of the hot and cold component;
$\mathrm{^g}$ redshift;
$\mathrm{^h}$ median values of the distribution of best-fit temperatures;
$\mathrm{^i}$ median values of the distribution of best-fit iron abundances. Lower and
upper errors correspond to the 16\% and 84\% percentiles, respectively. 
\end{list}
\end{table*}

\begin{figure}
\centering
\includegraphics[width=7.0 cm, angle=0]{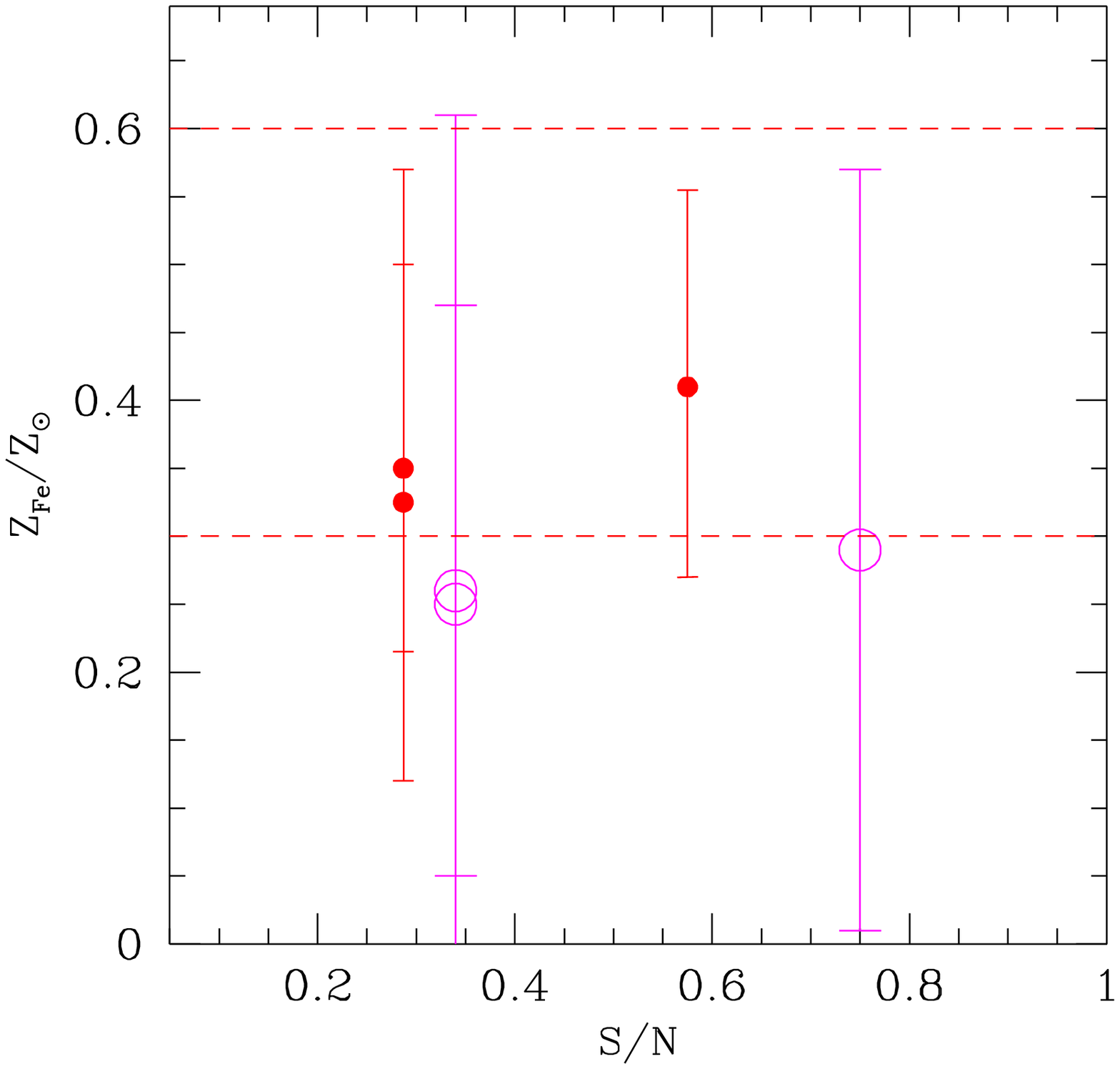}
\caption{Median of the best-fit $Z_{Fe}$ distribution as a function
of the ratio of the Cold and Hot Emission Measure, for simulations
where the cold component has an iron abundance twice higher than the
hot component ($0.6\,Z_\odot$ vs $0.3\,Z_\odot$). Filled circles are
for $z=0.6$, while empty circles are for $z=1$. Lower and upper error
bars correspond to the 16\% and the 84\% percentiles, respectively.}
\label{simul3}
\end{figure}

\subsection{Fitting bias in isothermal spectra rich in $\alpha$-element}

Finally, we checked whether a non-solar abundance ratio can
affect the measure of $Z_{Fe}$. In particular, we checked whether higher
abundances of S, Si and O can artificially yield a higher $Z_{Fe}$.
Therefore, we tried to recover the iron abundance with a {\tt mekal}
model, when the metallicity ratio among elements is higher than
solar.  The simulated spectra have the following input: $Z_{Fe} =
0.3$ and $Z_{\alpha} = 1$ or $Z_{\alpha} = 2$.  We recall that these
are extreme cases with respect to the abundances of $\alpha$-elements
observed in local X-ray clusters \citep[see][]{tam04}. We simulated
1000 spectra of a $z=0.4$ cluster with $kT = 3.5$~keV (2060 net
counts expected) and with $kT = 2$~keV (2300 net counts expected),
maximizing the presence of lines from $\alpha$-elements in our
energy range (lowest redshift and lowest temperature in our
sample). We find that the temperatures are slightly higher that the input
values in both cases ($kT = 3.8 \pm 0.29$ and $2.45 \pm 0.12$ keV), while
iron abundances are consistent ($0.30\pm 0.12\,Z_\odot$ and
$Z=0.25^{+0.13}_{-0.10}\,Z_\odot$ for $kT =3.5$ and 2~keV,
respectively). Therefore no bias is found in the measure of
$Z_{Fe}$, confirming the expectation that the iron abundance is
mostly determined by the K-shell complex at $6.7-6.9$ keV and is
not affected by the L-shell complex below 2 keV, where the iron
emission lines are blended to that of Si, O, and Mg.

\end{document}